\documentclass[twocolumn,twocolappendix]{aastex7} 
\usepackage{newtx}
\usepackage{float} 
\usepackage{longtable}
\usepackage{booktabs}
\usepackage{siunitx}
\usepackage{natbib}
\usepackage{caption}
\newcommand{\msun}{M$_\odot$}

\defcitealias{Toonen_2013}{TN13}

\begin{document}
\author[orcid=0009-0004-7915-2775,sname='Shani']{Y. M. Shani}
\affiliation{Department of Particle Physics and Astrophysics, Weizmann Institute of Science, Rehovot 7610001, Israel}
\email[show]{yarin-meir.shani@weizmann.ac.il}

\author[orcid=0000-0002-0430-7793,sname=Hallakoun]{Na\string'ama Hallakoun}
\affiliation{Department of Particle Physics and Astrophysics, Weizmann Institute of Science, Rehovot 7610001, Israel}
\email[]{} 

\author[orcid=0000-0001-6760-3074,sname=Ben-Ami]{Sagi Ben-Ami}
\affiliation{Department of Particle Physics and Astrophysics, Weizmann Institute of Science, Rehovot 7610001, Israel}
\email[]{} 

\author[orcid=0000-0001-9298-8068,sname=Shahaf]{Sahar Shahaf}
\affiliation{Max-Planck-Institut f\"ur Astronomie (MPIA), K\"onigstuhl 17, 69117 Heidelberg, Germany}
\affiliation{Department of Particle Physics and Astrophysics, Weizmann Institute of Science, Rehovot 7610001, Israel}
\email[]{} 

\author[orcid=0000-0002-3651-5482,sname=Li]{Jiadong Li}
\affiliation{Max-Planck-Institut f\"ur Astronomie (MPIA), K\"onigstuhl 17, 69117 Heidelberg, Germany}
\email[]{} 

\author[orcid=0000-0003-4996-9069,sname=Rix]{Hans-Walter Rix}
\affiliation{Max-Planck-Institut f\"ur Astronomie (MPIA), K\"onigstuhl 17, 69117 Heidelberg, Germany}
\email[]{} 

\author[orcid=0000-0002-2998-7940,sname=Toonen]{Silvia Toonen}
\affiliation{Anton Pannekoek Institute for Astronomy, University of Amsterdam, Science Park 904, 1098 XH Amsterdam, The Netherlands}
\email[]{} 

\title{The Silent Majority: The Interacting Post-Common-Envelope Binaries Underlying Cataclysmic Variables}
\shorttitle{Interacting Post-Common-Envelope Binaries}
\shortauthors{Shani et al.}

\begin{abstract}
We analyze the orbital period distribution of post-common-envelope white-dwarf-main-sequence (WDMS) binaries by cross-matching the new spectroscopic Gaia DR3 WDMS catalog with TESS light curves, and applying a uniform periodicity search and vetting pipeline. We identify 107 periodic systems, including 74 eclipsing binaries (32 new) and 33 binaries exhibiting only sinusoidal variations. Injection-recovery tests and a forward detectability model yield a completeness-corrected distribution that is well-described by a two-component function: a log-period Gaussian peaking at $P_{\rm orb} \approx 4.1 $\,h with $\sigma \approx 1.8$\,h, plus a rising component that begins near $P_{\rm orb}\approx12.9$\,h. We refer to this extended component as the long-period tail. It consists exclusively of detached non-interacting post-common-envelope binaries (PCEBs) that likely emerged from the common envelope and have not yet initiated mass transfer. In contrast, the short-period Gaussian is dominated by interacting or near-contact systems (including 22 known cataclysmic variables), consistent with high Roche-lobe filling factors. From the completeness-corrected distribution we infer that $29.8\%\pm4.5\%$ of the spatially unresolved WDMSs in our parent catalog are close PCEBs. Binary population synthesis models with high common-envelope efficiencies overproduce long-period systems and fail to reproduce the sharp peak, whereas lower efficiencies ($\alpha\lambda \leq 0.3$) match the peak more closely, yet still underpredict the tail. Our results hint at a large, currently under-classified reservoir of pre-cataclysmic variables and weakly accreting binaries, and provide new constraints on common-envelope physics.

\end{abstract}



\section{Introduction}
\label{sec:intro}

White dwarf-main sequence (WDMS) binaries play a crucial role in binary stellar evolution and the formation of compact objects \citep{Ivanova_2013}. Generally, WDMS systems comprise both wide, non-interacting, binaries and compact, post-common-envelope binaries (PCEBs). Close WDMS binaries serve as progenitors of interacting systems such as cataclysmic variables (CVs), AM CVn binaries, and double-degenerate binaries that may eventually trigger a Type Ia supernova \citep[\textit{e.g.}][]{Ivanova_2006}. These systems offer key insights into the physical processes governing binary interaction, mass transfer, and angular momentum loss.

The evolutionary phase shaping short-period WDMS binaries is most likely the common-envelope (CE) phase, during which the envelope of the evolving primary, the progenitor of the observed white dwarf (WD), engulfs the secondary. This process results in orbital decay and potential envelope ejection.

During the CE phase, drag forces within the envelope cause the orbit to contract, transferring orbital energy and angular momentum to the envelope, which may ultimately be ejected, forming a compact, short-period, PCEB containing a WD and a main-sequence (MS) star. In contrast to such systems, in cases where the envelope fails to eject, the stellar objects will merge. This process is often parameterized by the CE efficiency factor, $\alpha$ \citep{Zorotovic_2022}. This factor quantifies the effectiveness with which orbital energy is transferred to the envelope. Despite significant effort in developing the theoretical framework that governs envelope ejecta, this process, like other aspects of CE physics, is poorly understood. To obtain strong empirical constraints, well-characterized PCEB samples are necessary. These system parameters, such as period and mass distribution, are expected to carry signatures of the interaction between the ejected envelope and the binary orbit.

PCEB can lose additional orbital angular momentum over time due to magnetic braking and gravitational radiation. These will shrink PCEB orbits further, creating interacting systems.
CVs are a subgroup of such systems, in which a WD accretes matter from a companion star. These systems often exhibit a sudden significant increase in brightness as a result of accretion disk outbursts, nova explosions, and magnetic accretion variations (\textit{e.g.,} \citealt{Warner1995}). It is estimated that $40 - 50\%$ of PCEBs will evolve into CVs within a Hubble time \citep{schreiber_2003, zorotovic_2011}. If a substantial fraction of PCEBs indeed ends up as CVs, both period distributions should be consistent within the framework of binary evolution. 

The CV population exhibits two prominent features in its orbital period distribution. The first is a deficit of systems with orbital periods between 2 and 3\,h. This \textit{period gap} is associated with the cessation of magnetic braking. The second feature is a \textit{period minimum} around $80$\,min, below which CVs are rare (see figure~4 in \citealt{Donor_stars_analysis}). This minimum arises when the donor star becomes (partially) degenerate, causing its radius to increase as it loses mass. The Roche-lobe must then expand accordingly, resulting in an orbital period increase. Such systems are also referred to as \textit{Period Bouncers}. These properties provide stringent tests of Angular Momentum Loss (AML) models (via magnetic braking and gravitational radiation; \citealt{cv_evltn}), as well as donor star structure. 

Large-scale surveys, such as the Sloan Digital Sky Survey \citep[SDSS;][]{York_2000}, the Catalina Sky Survey \citep{Drake_2009}, and the Zwicky Transient Facility \citep[ZTF;][]{Bellm_2019}, have enabled the identification and study of several hundred PCEBs. In some cases, the binary plane of motion is sufficiently aligned with our line of sight, so that the two components are eclipsing. In these cases, we can obtain precise estimates of the orbital parameters \citep{SDSS_PCEBs,Catalina_PCEBs,Brown_2023}. Such surveys have provided valuable samples of eclipsing PCEBs and have developed methods to account for known selection effects. However, observational limitations such as sampling cadence, magnitude limits, and visibility constraints can still leave portions of the parameter space underexplored, particularly for short-period low-amplitude systems. Additional high-cadence observations from complementary facilities could help bridge these limitations and refine the measurements of occurrence rates and orbital period distributions.

The following study aims to empirically determine the intrinsic orbital period distribution of PCEBs. For this purpose, we compile a sample of eclipsing WDMS binaries by cross-matching the new spectroscopically identified Gaia XP WDMS catalog by \cite{Li2024} with NASA's Transiting Exoplanet Survey Satellite \citep[TESS;][]{TESS} photometry. We systematically search for periodic photometric variations in this sample and identify the periodicity for $\sim 107$ high-confidence WDMS systems. After accounting for selection effects and detection biases, we obtain a robust estimate for the intrinsic underlying PCEB period distribution. 

The structure of this paper is as follows. Section~\ref{sec:sample_selection} defines our sample and describes the main selection effects that influence it. Section~\ref{sec:periodicity_search} outlines the periodicity search in TESS light curves and the vetting procedures for eclipse detection. Section~\ref{sec:forward_modeling} introduces our forward-modeling approach for quantifying detection biases and deriving a completeness-corrected period distribution. In Section~\ref{sec:results}, we present both the observed and bias-corrected orbital period distributions. Section~\ref{sec:discussion} examines the implications for CE evolution and the formation of CVs. Finally, Section~\ref{sec:conclusions} summarizes our conclusions and outlines directions for future work.

\section{Sample Selection} \label{sec:sample_selection}

\subsection{WDMSs from Gaia XP spectra}\label{subsec:WDMS_from_Gaia_XP}

\begin{figure}
    \centering
    \includegraphics[width=1\columnwidth]{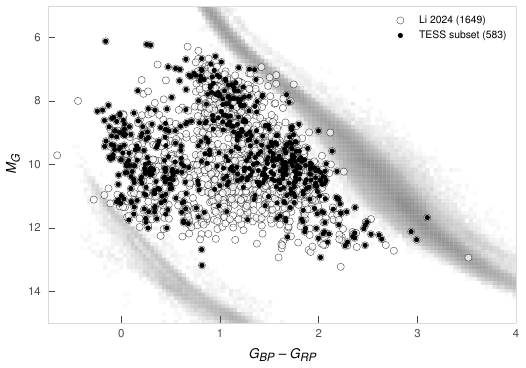}
    \caption{The location on the HR diagram of all sources in the \cite{Li2024} high-confidence WDMS sample (white circles) and those with available TESS data (black circles). The Gaia 100\,pc sample is shown as a gray two-dimensional histogram, for reference. WDMS candidates occupy a distinct locus between the MS and the WD sequence.}
    \label{fig:HRD}
\end{figure}

The catalog analyzed in this study is based on a sample of WDMS binary candidates identified by \citet{Li2024}. Their selection relied on low-resolution blue-photometer/red-photometer (BP/RP) spectra (also known as XP spectra) from Gaia's third data release (DR3). Using a model-independent neural network for spectral modeling and a Gaussian Process Classifier, they identified unresolved WDMS binaries among more than 10 million stars within 1\,kpc, yielding a sample of ${\sim}\,30{,}000$ candidates. A subset of this catalog was further validated through $\chi^2$-based spectral fitting with composite binary models, resulting in 1,649 high-confidence WDMS binaries, which form the initial sample for this study. Figure~\ref{fig:HRD} shows the initial WDMS sample on the Hertzsprung-Russell (HR) diagram, highlighting their location between the MS and the WD sequence, consistent with significant flux contributions from both stars. 

Cooling ages and masses for the sample were derived with the \texttt{WD\_models} package\footnote{\url{https://github.com/SihaoCheng/WD_models}}, assuming C/O-core WDs with thick hydrogen atmospheres \citep{bedard_2020}. Central masses were obtained by converting the best-fit photometric parameters $G_\text{BP}-G_\text{RP}$ and $M_G$ of the WD component from the XP-spectrum disentangling \citep{Li2024}, to mass using the WD evolutionary grids. Uncertainties were estimated by Monte Carlo resampling: 1000 realizations of $G_\text{BP}-G_\text{RP}$ and $M_G$ were drawn from normal distributions centered on the measured values; each realization was converted to mass and cooling age, and asymmetric errors were taken from the 16th, 50th, and 84th percentiles of the resulting distributions. Typical uncertainties are $\approx 0.3\,M_\odot$ for the mass, and $\approx 0.1 $\,Gyr for the cooling age. This procedure propagates photometric errors into both the mass and cooling-age estimates.

\subsection{TESS Light Curves}
\label{sec:tess_lcs}

To identify eclipsing binaries (EBs) within our WDMS sample, we used photometric data from TESS. The satellite observes the sky in overlapping ${\sim}\,24^\circ\times \ 96^\circ$ sectors, each continuously monitored for ${\sim}\,27$~days, providing light curves at multiple cadences: 20\,s and 2\,min for selected targets, and 200\,s, 10\,min and 30\,min from full frame images (FFIs). 

We cross-matched the Gaia DR3 identifiers of 1,649 WDMS binaries with the TESS Input Catalog \citep[TIC;][]{TIC} using the Mikulski Archive for Space Telescopes (MAST)\footnote{\url{https://mast.stsci.edu}}, identifying 583 systems with publicly available TESS time-series data. For each matched source, we used the \texttt{Lightkurve}\footnote{\url{https://docs.lightkurve.org}} \textsc{Python} package to retrieve its sector-level light curves. A subset of the Jiadong-TESS cross-matched systems and their reported parameters is shown in Table~\ref{tab:sample}.

The collected light curves were produced from the four primary TESS pipelines: (i) \texttt{SPOC} \citep{Jenkins_2016}, which produces calibrated short-cadence data for pre-selected targets; (ii) \texttt{TESS-SPOC HLSP} \citep{Caldwell_2020}, which extends SPOC-style photometry to FFIs for up to 160,000 field stars per sector; (iii) \texttt{Quick-Look Pipeline} \citep[\texttt{QLP};][]{Haung_2020}, which processes FFIs for stars with $T < 13.5$\,mag using multi-aperture photometry (where $T$ is the TESS passband); (iv) \texttt{GSFC-ELEANOR-LITE} \citep{eleanor}, which provides 30-min photometry optimized for targets with $T < 16$\,mag in crowded fields; And (v) the \texttt{TESS-Gaia Light Curve} \citep[\texttt{TGLC};][]{tglc}, which delivers PSF-based light curves from FFIs using Gaia DR3 as priors, forward modeling effective PSF for decontamination.

The final sample comprises 307 systems with \texttt{SPOC} light curves, 46 additional sources from \texttt{TESS-SPOC HLSP}, 180 from \texttt{GSFC-ELEANOR-LITE}, 46 from \texttt{QLP}, and 4 from \texttt{TGLC}, which are not covered by the other pipelines.

Most light curves include both simple aperture photometry (SAP) and systematics-corrected flux (Pre-search Data Conditioning Simple Aperture Photometry; PDCSAP), the latter derived using co-trending basis vectors. We consistently adopted the PDCSAP flux where available and used only data points flagged as high quality (\texttt{QUALITY} = 0).

To avoid introducing stitching artifacts or cross-sector inconsistencies, we analyzed each sector independently.

\subsection{Selection effects}
\label{subSec:selection_effects}
Interpreting any observed period distribution requires understanding how the sample differs from a volume-complete reference population. These differences arise at three stages: (i) the construction of the \citet{Li2024} \textit{high-confidence} Gaia-XP catalog, (ii) the availability of TESS data for the systems, and (iii) the geometric and signal-to-noise ratio (SNR) constraints on eclipse detection. In summary, the final sample we analyzed is comprised of systems that are likely the product of a CE phase. The sample is biased toward short-period, unresolved, young WDMS binaries with relatively low masses. The biases are the result of catalog construction, TESS target selection, and the geometry of eclipse detection, all of which we detail below. The analysis described in later sections takes these effects into account. 

\subsubsection{Catalog completeness (Gaia XP)}\label{subsec:Catalog_Cuts}

The \citet{Li2024} catalog identifies WDMS binaries by detecting both the WD and the companion spectral features in Gaia XP spectra. This introduces selection effects that become increasingly important with distance. At the catalog's distance limit of 1\,kpc, the SNR threshold corresponds to an absolute magnitude limit of $M_{G} \simeq 14$. As a result, older, cooler WDs (with cooling ages $>3\,$Gyr) often fall below this detection threshold and are excluded.

 Additionally, the catalog is optimized for systems in which the WD and the MS contribute comparable optical flux ($|M_{G\text{,MS}} - M_{G\text{,WD}}| < 2.5$). Consequently, our starting sample underrepresents systems where: (a) hot WDs are paired with late MS types, leaving no detectable imprint, (b) cool WDs orbit bright MS companions, and (c) the binary orbit is wide enough to be spatially resolved by Gaia.

Additional selection criteria (an apparent magnitude of $G \lesssim 20.7$, and extinction \(A_G \lesssim 0.5\)) further bias the sample towards high-latitude fields and against dusty regions near the Galactic plane, where extinction and crowding affect Gaia. Additionally, the low \texttt{RUWE} criterion (\texttt{RUWE}\,$<1.4$) disfavors systems with orbital periods longer than $\mathcal{O}(10)$ of days.

\begin{figure}
    \centering
    \includegraphics[width=1\linewidth]{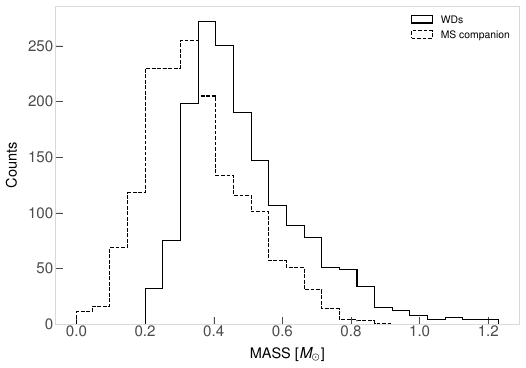}
    \caption{Mass distributions of the WDs (solid histogram) and MS companions (dashed) in our WDMS sample.
WD masses concentrate at $0.3-0.5$\,M$_\odot$, while companions are mostly $0.075$-$0.6$\,M$_\odot$ (late-type M dwarfs).
}
 
    \label{fig:mass_distribution}
\end{figure}

These selection biases inherently favor systems with relatively low-mass WDs, which are larger and typically brighter at a given cooling age. Such WDs are more likely to originate from binaries that underwent an early CE evolution, which has been shown to produce tighter binaries \citep{scherbak_2022}. Furthermore, we expect the companions to be late-type stars (typically M dwarfs), so the WD's spectral signature can be resolved. Indeed, we find that the WD mass distribution is skewed towards smaller values, and that the companions are mostly M-dwarfs. As noted in Section~\ref{subsec:WDMS_from_Gaia_XP}, the typical uncertainty on an individual WD mass is $\approx 0.3\,M_\odot$, which broadens the observed distribution; however, in the absence of evidence for systematics, these errors are not expected to shift the distribution's central tendency.
Indeed, we find that the WD-mass distribution is skewed toward smaller values and peaks near $\approx 0.4\,M_\odot$, while the companions are mostly M dwarfs. This is reflected in the mass distribution shown in Figure~\ref{fig:mass_distribution}, where the WD mass distribution (solid outline) peaks at $0.4$\,M$_\odot$, while the companion masses (dashed outline) mostly fall in the $0.075-0.6$\,M$_\odot$ range, as expected for late-type M dwarfs.

\subsubsection{TESS target selection bias}
TESS short-cadence observations (2\,min and 20\,s) are limited to a subset of pre-selected bright stars ($T_{\text{mag}}\lesssim 15$), prioritized for planet detection, resulting in a known observational bias toward optically bright and nearby sources \citep{TIC}, and against intrinsically faint WDMS binaries, even if present in the Gaia catalog. 
By incorporating TESS's FFIs and long-cadence light curves from various pipelines, we increase our available sample to include fainter systems, extending practical coverage to TESS's limiting magnitude of $\sim17$\,mag.

To assess potential observational biases introduced by TESS selection functions, we performed two-sample Kolmogorov-Smirnov (KS) tests independently on each axis of the HR diagram, comparing absolute magnitude and color distributions between the full sample and the subset available in TESS. The tests do not indicate significant differences in the absolute magnitude distribution ($p \approx 0.67$). However, there is a significant difference in the color distribution ($ p \approx 5.7 \times 10^{-6}$), suggesting a notable color-based selection bias in the TESS observations. The TESS mission uses a red-optical bandpass covering the wavelength range from about 600 to 1000\,nm, making it particularly sensitive to cooler, red stars, and therefore creating an observational bias towards this stellar population.

\subsubsection{Geometric bias}

Eclipses require a nearly edge-on system's inclination. For a circular orbit with an impact parameter $b = 0$, the probability of an eclipse along a random line of sight is
\begin{equation} \label{eq:geometric_probability}
    p_\text{geo} = \frac{R_1+R_2}{a},
\end{equation}
where $a$ is the orbital separation, and $R_1\,,R_2$ are the stellar radii.
Short orbital periods correspond to smaller separations so that $p_{\rm geo} \propto P^{-2/3}$ for a given system total mass, which biases an eclipsing sample against longer periods. Additionally, eclipse detection requires sufficient depth relative to the photometric precision. Shallow eclipses in low SNR light curves may go undetected. We model these detection biases in detail in Section~\ref{sec:forward_modeling}.

\section{Periodicity search}\label{sec:periodicity_search}

To identify eclipsing WDMS systems, we performed a systematic search for periodic signals in the TESS light curves using the fast Box-Least Squares (fBLS) algorithm \citep{Shahaf_2022}, which is optimized for the detection of box-shaped signatures. While we expect many WDMS binaries to show sharp ingress and egress due to their short orbital periods and to exhibit flat-bottomed eclipses due to the typical radius ratio, resulting in box-like signals, the fBLS algorithm remains sensitive to a broader range of eclipse morphologies, including grazing eclipses, as long as periodic dips are present. Additionally, it is capable of detecting other types of periodically variable systems, such as those exhibiting reflection effects or ellipsoidal variations, which may produce sinusoidal flux modulations rather than distinct dips.

\subsection{Period Analysis and Eclipse Identification}

We computed the fBLS periodogram of each light curve (for each combination of sector and cadence individually) over a uniformly spaced frequency grid, corresponding to orbital periods ranging from 1\,h to 10~days. 
The lower bound reflects the Nyquist limit for a 30-min cadence and is below the theoretical minimum period for CVs ($\sim$80\,min). The maximum detectable period of 10~days ensures that at least two full orbital cycles are captured within a single TESS sector observation, the minimum requirement for reliable periodic signal identification. We chose not to search beyond the duration of a single sector. This scanning range comfortably covers the expected period regime of PCEBs and does not favor targets with more available observations. Furthermore, the probability of observing an eclipse decreases with the period, hinting that we might miss $\mathcal{O}(1)$ systems.

Each fBLS periodogram consists of scores for $\sim 150,000$ frequencies and is generated by folding the light curve at each trial period and evaluating a box-shaped model using a 200-bin fast-fold profile. We allow significant eclipse widths to allow detection of both detached and contact binaries.

The fBLS algorithm outputs a detection score (approximation of the BLS score) for each trial period, along with the width and bin configuration. The best-fit period for each light curve was selected as the one with the highest fBLS score. 

We commonly define the SNR as the eclipse depth divided by the error on the depth, as the fBLS assumes box-car transits. For Poisson noise statistics and noise per unit time $\sigma_0$, the noise over an interval of the transit duration will be $\sigma = \sigma_0 / \sqrt{T_\text{eclipse}}$, and therefore,
 \begin{equation}
     SNR = \frac{\delta}{\sigma_0}\sqrt{T_\text{eclipse}},
 \end{equation}
 where $\delta$ is the normalized transit depth. Noting that, as one phase-folds $N$ transits together, then $\sigma_0 \rightarrow \sigma_0/\sqrt{N}$ \citep{Kipping_2023}.
 
We computed the eclipse SNR for every folded binned light curve and required a value larger than 5 as a threshold for detection. At the end of this periodicity search, we compiled a list of eclipsing candidates with orbital periods ranging from $\sim1.4$\,h up to $\sim6$~days. These candidates then underwent a thorough validation process as described below.

\subsection{Detection Validation} \label{SubSection:detection_validation}

We subjected each eclipse candidate to a series of diagnostic tests to confirm its astrophysical nature and to rule out false positives (such as an eclipse of a neighboring star or a transiting exoplanet scenario). Our validation procedure was inspired by the Data Validation Reports (DVR) provided by \texttt{SPOC} and \texttt{QLP} and the frameworks used in exoplanet transit vetting (e.g., the \texttt{LATTE} light curve analysis tool; \cite{Eisner2020}) and by the TESS EB survey of sectors 1--26 \citep{Prsa_2022}, which utilized \texttt{ICED LATTE} for EB validation.

Certain diagnostics can provide an unambiguous confirmation. An existing SIMBAD classification as an EB or a CV, or a pronounced, on-target secondary eclipse whose depth cannot be reproduced by diluting neighbors, secures the EB status. When a single decisive indicator is absent, each diagnostic contributes a pass or fail flag. The resulting pattern is carried into the catalog. None of the candidates is marked as a TESS Object of Interest, confirming there are no planetary candidates in the sample. In brief, we performed the following checks for each candidate.

\subsubsection{Pixel-level localization}
For each source with an available calibrated Target Pixel File (TPF), the light curve of each pixel was phase-folded and its eclipse SNR computed. The pixel with the highest SNR must coincide with or be adjacent to the catalog position of the target. A mismatch flags the event as \texttt{OffPix}. A complementary in- and out-of-transit difference image was generated and expected to present a flux deficit peak in the same pixel's neighborhood. The image was stored in case a visual inspection is required.

\subsubsection{Archival consistency}
A SIMBAD object query by Gaia source ID and an inspection of the \texttt{SPOC} data-validation (DV) report were performed. A positive EB/CV classification or a suspected EB flag in the DV report accepted the target (\texttt{ArchivalOk}) without further testing.

\subsubsection{Dilution assessment}
Gaia DR3 neighbors within $60''$ and up to five magnitudes fainter were modeled with a $20''$ Gaussian point-spread function (PSF) to estimate the contamination fraction. All systems with brighter neighbors received a \texttt{BrightNbr} flag.

\subsubsection{Secondary-eclipse search}
At phase 0.5, a box of equal duration to the primary was fitted. A detection with SNR $\geq 3 $ and depth $\geq 10 \%$ of the primary supports the EB hypothesis. Null detection was labeled as \texttt{SecMiss}.

The validation process is summarized in a bit-mask column \texttt{VET\_FLAGS}, which records any failed checks: \texttt{OffPix}, \texttt{BrightNbr}, \texttt{SecMiss}. Flagged entries remain in the catalog, but are highlighted for caution.

We omit five eclipsing detections as their pixel-level analysis was inconclusive, as a bright neighbor was imaged on the same pixel as the candidate. A couple of initially flagged cases with high-contamination sources were anticipated. Overall, we found that 107 of the candidates passed the validation tests, with 74 presenting clear eclipses, and the remaining 33 present sinusoidal variations. We list the eclipsing sample Gaia source IDs, coordinates, Gaia magnitudes, orbital periods, and classification, if they exist, in Table~\ref{tab:detected_systems}. Figures~\ref{fig:ecl1}~-~\ref{fig:sinusoidal} show representative phase-folded and binned light curves for our detected sample: CVs (Figure~\ref{fig:ecl1}), EBs and unclassified eclipsing systems (Figure~\ref{fig:ecl2}), and sinusoidal variables (Figure~\ref{fig:sinusoidal}).

\begin{figure}
    \centering
    \includegraphics[width=1\columnwidth]{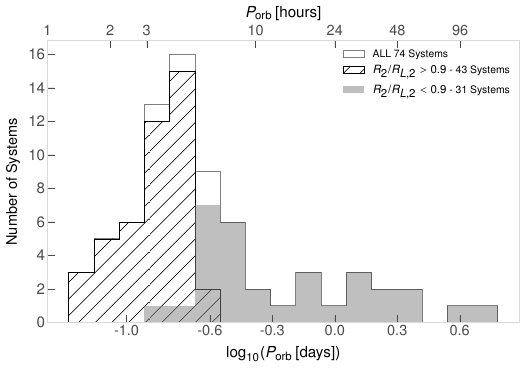}
    \caption{Observed orbital period distribution for the 74 validated eclipsing WDMS binaries (CVs included). The distribution shows a clear peak near $\sim$4\,h and a tail extending to a few days. Systems are color-coded by their Roche-lobe fill factor (see Appendix~\ref{sec:roche_method}): near-contact systems with fill factor $\geq 0.9$ \emph{(hatched histogram)} , and detached systems with fill factor $< 0.9$ \emph{(filled gray)}. This separation highlights the connection between orbital period and binary interaction state. We label fill-factor $\geq 0.9$ systems as interacting PCEBs which may include pre-CVs and CVs, and $<0.9$ as detached PCEBs.
}
    \label{fig:period_dist_fill_factor}
\end{figure}

In the following section, we analyze the orbital period distribution of this validated sample while accounting for detection biases. Figure~\ref{fig:period_dist_fill_factor} shows the observed period distribution for 74 eclipsing systems, featuring a prominent peak around $\sim$4\,h and a residual tail extending to longer periods of up to a few days. This motivates us to further subdivide the distribution based on the inferred Roche-lobe fill factor (see Appendix~\ref{sec:roche_method}) using the provided mass estimation of the binary, separating near-contact binaries (fill factor $\geq 0.9$) from detached systems. The clear contrast between the two subpopulations underscores the role of binary interaction in shaping the observed period distribution.

\section{Forward modeling of the period distribution} \label{sec:forward_modeling}

The observed period distribution of our eclipsing sample is shaped by the selection effects discussed in Section~\ref{subSec:selection_effects}. In short, the probability of detecting an eclipse decreases with increasing orbital separation, introducing a geometric bias that favors short-period systems. The finite TESS observing windows, cadence, and detection threshold cause binaries with longer periods, shallower eclipses, or unfavorable eclipse timings to be systematically missed. Additionally, the sensitivity of our detection pipeline further shapes the systems that appear in the final observed distribution.

To recover the intrinsic orbital period distribution of PCEBs, we correct for these biases. We approach this by constructing a model for the detection probability as a function of period, $p_{\rm det}(P_{\rm orb})$, and then applying it to invert the observed distribution.

We begin by considering the total parameter space of the WDMS binaries. The underlying population can be formally described by a joint probability distribution,
\begin{equation}
    p(M_\text{MS},M_\text{WD},P_\text{orb},e,i),
\end{equation}
where $M_\text{MS}$ and $M_\text{WD}$ are the mass of the stellar components, $P_\text{orb}$ is the orbital period, $e$ is the eccentricity, and $i$ is the orbital inclination. We assume that the inclination is statistically independent of other properties and that the eccentricity can be neglected for close PCEBs, which are expected to be circularized. This allows us to factor the distribution as follows:
\begin{equation}
    \begin{split}
    p(M_\text{MS},M_\text{WD},P_\text{orb},e,i) & \approx \\
    p(M_\text{MS},M_\text{WD}) & \times p(P_\text{orb}|M_\text{MS},M_\text{WD})\times p(i).
    \end{split}
\end{equation}

Although we do not explicitly simulate this population, this factorization motivates our forward-modeling approach. To recover the intrinsic distribution of orbital periods, $p(P_\text{orb})$, from the observed distribution, $\hat{p}(P_\text{orb})$, we account for selection effects that filter the true population into the detected sample. We therefore model the observed period distribution as,
\begin{equation}
    \hat{p}(P_\text{orb}) = p(P_\text{orb})\times p_\text{det}(P_\text{orb}),
\end{equation}
where $p_\text{det}(P_\text{orb})$ is the probability of detecting a system with period $P_\text{orb}$ as eclipsing in our search.

To model the detection probability, we performed an injection-recovery analysis based on a synthetic population drawn from binary population synthesis (BPS; see Section~\ref{sec:BPS} below). We applied the same photometric and mass-based selection criteria used for the \citet{Li2024} catalog (i.e., color contrast, secondary mass, and Gaia color-magnitude cuts. See Section~\ref{subsec:Catalog_Cuts}) to define a subsample suitable for light-curve simulation. From this filtered population, we sampled systems by drawing their periods, masses, radii, and effective temperatures from the empirical distributions. 

Each system was assigned with a random TESS light curve from one of the pipelines discussed in Section~\ref{sec:tess_lcs}, chosen solely for its cadence and photometric scatter. The modeled light curves were generated using the Physics Of Eclipsing Binaries (\texttt{PHOEBE}) v2.3 \textsc{Python} package \citep{Prsa2016}, assuming Roche geometry, blackbody atmospheres, zero limb-darkening coefficients, and orbital inclinations uniformly distributed in $\cos i$ to factor in geometric selection effects. Systems exceeding their Roche limits were treated via semi-detached or contact binary configurations, using \texttt{PHOEBE}'s geometry constraints. 

We estimated the photometric precision of each TESS light curve by computing its Combined Differential Photometric Precision (CDPP), evaluated at the native cadence of the data. The resulting $1\sigma$ white noise level is then added to the paired model flux to simulate realistic TESS photometric conditions. In the same detection routine, we defined a match when the recovered period agreed with the injected period (or one of its harmonics) within 5$\%$ and SNR $\geq 5$. The efficiency curve was calculated in uniform bins of logarithmic $P$ hours.

The results of these injections are a set of detection efficiency curves $\epsilon_i(P_\text{orb})$ for each cadence class. These are averaged to derive a mean detection efficiency $\epsilon(P_{\text{orb}}) \equiv p(P_{\rm orb})$.

Acquiring $\epsilon(P_{\text{orb}})$ allowed us to form a global completeness function, which reflects the typical detectability of eclipses as a function of the orbital period. To quantify the diversity in detection sensitivity across the population, we propagated the $\pm 1 \sigma$ spread in detection probability at each period. This envelope captures the variance in eclipse visibility due to differences in stellar radii, orbital separation, and photometric depth. The final completeness curve is thus a function of the orbital period alone, but takes into account the physical diversity and photometric detectability in our sample.

\begin{figure}
    \centering
    \includegraphics[width=\columnwidth]{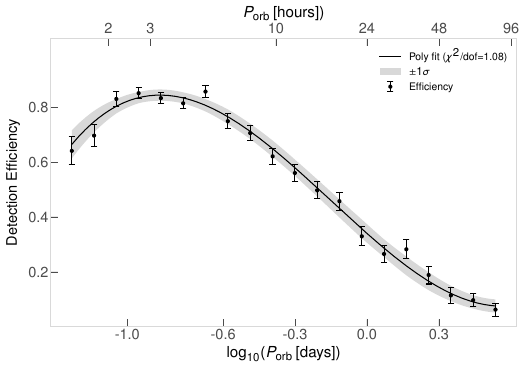 }
    \caption{Mean eclipse detection probability as a function of orbital period. The black curve shows the average detection efficiency across all eclipsing systems in the sample, incorporating geometric alignment and photometric detectability. The gray-shaded region denotes the $\pm 1 \sigma$ spread in detection probability, reflecting population-level variance in system properties such as stellar radii and flux ratio. This completeness curve is used to correct the observed period distribution and recover the intrinsic distribution.}
    \label{fig:completness_curve}
\end{figure}

Figure~\ref{fig:completness_curve} plots the resulting completeness curve (black) with the gray shaded region for the 1$\sigma$ spread. As expected, detection efficiency is highest at short orbital periods, where eclipse probability is most favorable and multiple eclipses are present, and declines toward longer periods. The decline towards the short-period end is a result of the decreasing efficiency of the 30-min cadence class. 

The efficiency curve derived from our injection analysis represents the combined probability that a WDMS binary will eclipse and be successfully identified by our detection pipeline. This includes the effects of geometric alignment, photometric detectability, and survey cadence. We find that the total detection efficiency remains higher than the naive geometric eclipse probability (Eq.~\ref{eq:geometric_probability}). These differences are mostly attributed to the sensitivity of our pipeline to out-of-eclipse modulations such as ellipsoidal variations, which are modeled in detail with \texttt{PHOEBE}. As a result, our completeness estimates capture not just idealized eclipses, but the full range of detectable photometric signatures of compact binaries.

Combining all these elements, the expected number of detected systems in a period bin is
\begin{equation}
    \hat{N}(P_\text{orb}) = f(P_\text{orb})\times p_\text{det}(P_\text{orb})\times N_\text{sample},
\end{equation}
where $f(P_\text{orb})$ is the intrinsic normalized period distribution of PCEBs, such that $\Sigma f(P_\text{orb}) = 1$, and $N_\text{sample}$ is the total number of systems in our sample.

\section{Results}\label{sec:results}
\subsection{Period Distribution}
We determined orbital periods for 74 eclipsing systems in our sample. All periods were derived using the same detection pipeline. The observed period distribution spans from 1.26\,h ($\sim$75\,min) to approximately 145\,h ($\sim$6\,d), and presents a quasi-normal distribution in $\log(P_{\rm orb})$, which peaks at roughly 4\,h with a spread of $\sim0.25$\,dex in $\log(P_{\rm orb}[\text{h}])$, along with a non-negligible population of systems extending into longer periods.

\begin{figure}
    \centering
    \includegraphics[width=1\columnwidth]{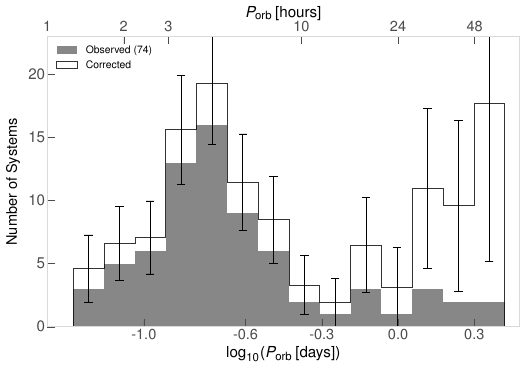}
    \caption{Orbital period distribution of the 74 PCEBs discovered as eclipsing in TESS data (gray), the bias-corrected distribution (white) and its errors.} 
    \label{fig:corrected_distribution}
\end{figure}

To recover the intrinsic orbital period distribution, we correct for biases by inverting the detection efficiency function derived in Section~\ref{sec:forward_modeling}. Figure~\ref{fig:corrected_distribution} compares the observed and bias-corrected distributions.

To model the corrected orbital period distribution of WDMS binaries, we adopt a two-component mixture model, a Gaussian in log-period to describe the primary short-period population, and a rising linear tail to capture the extended distribution at longer periods. Specifically, we fit the function
\begin{equation}
\begin{aligned}
& f(\log P) =  w \cdot \mathcal{N}(\mu, \sigma)  \\ & +  (1-w) \cdot m \cdot (\log P - \log P_{\mathrm{break}}) \cdot \Theta(\log P - \log P_{\mathrm{break}})
\end{aligned}
\end{equation}
where $\mathcal{N}(\mu, \sigma)$ is a Gaussian in $\log(P_{\mathrm{orb}}[h])$, and the second component is a rising linear function that starts at a break point $\log P_{\mathrm{break}}$. The slope $m$ and break location $P_{\mathrm{break}}$ were selected via a grid search to minimize the $\chi^2$ statistic relative to the corrected histogram, and for each pair of $(m, P_{\mathrm{break}})$ we optimized the Gaussian parameters $(\mu, \sigma, w)$ using maximum likelihood estimation.

To avoid biases introduced by poorly constrained bins, we excluded the last three period bins (beyond $\sim$72\,h) from the fit. These bins are subject to large completeness corrections due to observational limits, and their inclusion would risk overfitting uncertain regions of the distribution.

\begin{figure}
    \centering
    \includegraphics[width=1\columnwidth]{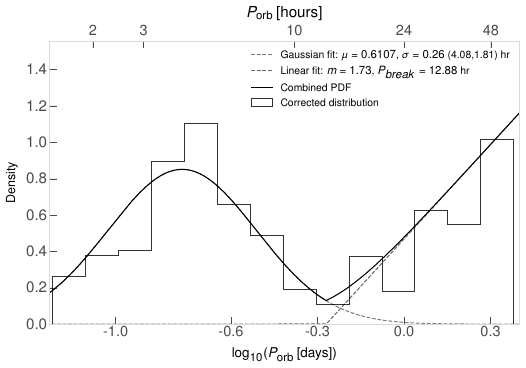}
    \caption{
    Mixture model fit to the corrected orbital period distribution. The best-fit model (solid black line) combines a Gaussian peak (dashed gray) with a rising linear tail (dashed gray) beginning at $\log_{10} P_{\mathrm{break}} = 1.1 \pm 0.12$ dex ($\approx$12.9\,h). The Gaussian component peaks at $\mu = 0.61 \pm 0.03$ dex ($\approx$4\,h) with a width of $\sigma = 0.26 \pm 0.07$ dex. The slope of the linear component is $m = 1.7 \pm 0.9$. Histogram bins represent the completeness-corrected sample, the last three bins (at $P_\text{orb} \gtrsim 72$\,h) were excluded from the fit due to large correction factors.
    } 
    \label{fig:gauss_linear_fit}
\end{figure}

The best-fit model is shown in Figure~\ref{fig:gauss_linear_fit}. We find $\mu = 0.61 \pm 0.03$ dex, corresponding to $P_{\mathrm{peak}} \approx 4$\,h, and $\sigma = 0.26 \pm 0.07$ dex. The linear tail begins at $\log_{10} P_{\mathrm{break}} = 1.1 \pm 0.12$ dex (i.e., $P_{\mathrm{break}} \approx 13$\,h) with a slope of $m = 1.7 \pm 0.9$. This structure supports a dual-regime interpretation for WDMS binaries, a dominant short-period peak, and a broad, shallow long-period tail.

A natural physical interpretation emerges when separating the population by Roche-lobe fill factor (see Figure~\ref{fig:period_dist_fill_factor}). Systems within the range of the Gaussian distribution are close to Roche-lobe filling (fill factor $\geq0.9$), indicating interacting or near-interacting PCEBs. 
Extending from the shortest detected periods up to the break near $P_{\mathrm{break}} \sim 13$\,h, this component is best interpreted as the interacting population of PCEBs. This regime includes the 22 classified CVs in our sample and likely many additional pre-CVs and weakly accreting systems whose signatures are too faint for CV classification, or just unclassified systems. In contrast, the long-period tail beyond $P_{\mathrm{break}}$ consists exclusively of detached PCEBs that emerged from the CE at wider separations and have not yet begun interaction. Thus, the corrected distribution naturally divides into two regimes: an interacting population represented by the Gaussian peak (CVs and pre-CVs, or interacting-PCEBs), and a detached population at longer periods representing the non-interacting PCEBs. In other words, the peak captures the pathway into CV evolution, while the tail reflects the detached population. We use post-CE WDMS for binary systems that have undergone a CE phase. We reserve detached PCEB for non-interacting systems, and interacting PCEB for systems with ongoing or imminent mass transfer (pre-CVs, weakly accreting binaries, and CVs). Accordingly, we refer to the Gaussian peak as the interacting post-CE component and to the long-period tail as the detached PCEB component.

This dual-regime structure suggests that the present-day PCEB population is not monolithic but shaped by a combination of CE physics and post-CE orbital evolution. The slope of the tail (m $\approx 1.7$) implies an increase in system numbers toward longer periods, once detectability biases are removed, though our constraints beyond $\sim 3$\,d remain limited by small-number statistics. A larger, deeper sample will be essential to confirm whether this feature persists and to test whether it reflects multiple CE efficiency regimes, mass-ratio dependent envelope ejection, or observational incompleteness.

\section{Discussion}\label{sec:discussion}

We identified 74 eclipsing WDMS binaries in TESS data, of which 22 are previously known CVs and 18 are known EBs, revealing 32 new eclipsing WDMS systems. And confirming two systems (TIC\,231050794 and TIC\,375745959), which carry an EB flag in TESS DVRs.

The orbital period distribution of this sample shows a clear log-normal-like peak centered near $P_{\mathrm{orb}} \sim 4$\,h, and an extended tail reaching several days. After correcting for detection biases, the population naturally separates into two regimes, a short-period component dominated by interacting systems (CVs, pre-CVs, and weakly accreting binaries), and a long-period component representing detached PCEBs that emerged from the CE at wider separations.

The overall distribution, with the majority of systems concentrated at orbital periods of a few to several hours, supports the picture that CE evolution typically produces very compact binaries. Such short orbits are a natural outcome of orbital energy and angular momentum loss during the envelope spiral-in, and their prevalence places useful constraints on CE physics \citep{1976Paczynski,Ivanova_2013}. While we argue that many of the shortest-period systems are interacting PCEBs (CVs, pre-CVs, or weakly accreting binaries), and the longer-period systems are more likely detached, we emphasize that a strict division between these regimes requires further spectroscopic confirmation of accretion signatures. Nonetheless, the existence of a tightly clustered short-period population alongside an extended long-period component suggests a more complex structure than current BPS models reproduce (see Section~\ref{sec:BPS}), while keeping in mind our corrected distribution might still include an unclassified CV contribution at the shortest periods. In particular, even though none of the tested BPS models match the observed distribution exactly, the relatively better agreement with the inefficient CE model ($\alpha\lambda \sim 0.25$) suggests that our results may help constrain the efficiency of envelope ejection.

\subsection{Comparison to previous work}
Our completness-corrected period distribution Gaussian component peaks at $\sim$4\,h with $\sigma_{\rm logP} \approx 0.25$\,dex ($\sim 1.8$\,h)  notably shorter and narrower than the RV-selected SDSS PCEB distribution of \cite{SDSS_PCEBs}, which spans 1.9\,h $-$ 4.3\,d and is well described by a log-normal with peak $P_{\rm orb} \approx 10$\,h and width $\sigma_{\rm logP} \approx 0.41$\,dex. This contrast is consistent with selection effects.
For the Catalina survey (CRTS), \cite{Catalina_PCEBs} searched 835 spectroscopically confirmed SDSS WDMS systems and found 29 eclipsing PCEBs (17 known, 12 new), noting that eclipse selection naturally skews toward short periods. A subsequent analysis \citep{Parsons_2015} added 14 eclipsing WDMS and many ellipsoidal systems whose M2–M3 donors place them tightly around $\approx4$\,h, broadly consistent with our hour-scale peak and decline toward longer periods.
For the ZTF survey, \cite{Keller_2021} cross-matched Gaia WD candidates with ZTF DR3 light curves and identified 18 new WD binaries (17 eclipsing) with periods 1.03$-$ 5.72\,h, emphasizing short-period sensitivity from eclipse geometry and cadence, while \cite{LiAndZhang_2024} analyzed $\sim200$ emission-selected WDMS and found 55 PCEBs spanning 2.2 $–$ 81.6\,h (six new, three eclipsing), thereby recovering more intermediate periods than eclipse-selected searches.
Finally, comparison to the Potential Post-Common Envelope Binaries catalog of \cite{Kruckow_2021} using the subset with $M_2 \leq 0.8$\,M$_\odot$ ($N=53$) shows a period distribution consistent with ours, with eight systems in common. Full per-survey details, figures, and a summary table are provided in Appendix~\ref{sec:comparison_to_previous_work}.

\subsection{CVs within the sample}
In our sample, 22 systems have been previously classified as CVs, comprising almost a third of the eclipsing PCEBs sample. While many CVs are expected to be post-CE, viable triple-induced channels (\textit{e.g.}, eccentric Kozai-Lidov excitation) can bring detached WDMS binaries into contact without a prior CE phase (see \citet{Knigge_2022,shariat_2025}). Their presence highlights that the shortest-period end of our distribution is already populated by interacting systems. The cutoff below $\sim 1.5$\,h aligns with expectations for CV evolution, and the absence of a $2-3$\,h ``period gap'' in the combined PCEB distribution reflects the fact that detached binaries evolve smoothly across this range. When the CVs are considered separately, the canonical period gap reappears (3 CVs are within $\leq 15\%$), confirming that it is a feature of mass-transferring systems rather than of the underlying post-CE population.

This fraction of CVs provides a firm lower limit on the proportion of PCEBs that will evolve into interacting binaries. In practice, the true fraction is likely higher, as many of the short-period systems without a CV classification may in fact be weakly accreting or pre-CV binaries. Thus, our results strengthen the view that the observed short-period peak of the PCEB distribution is dominated by the pre-CVs, while the long-period tail is shaped by detached, non-interacting systems.

\subsection{Predicted vs. Observed Period Distributions}
\label{sec:BPS}

\begin{figure*}
    \centering
    \includegraphics[width=0.9\textwidth]{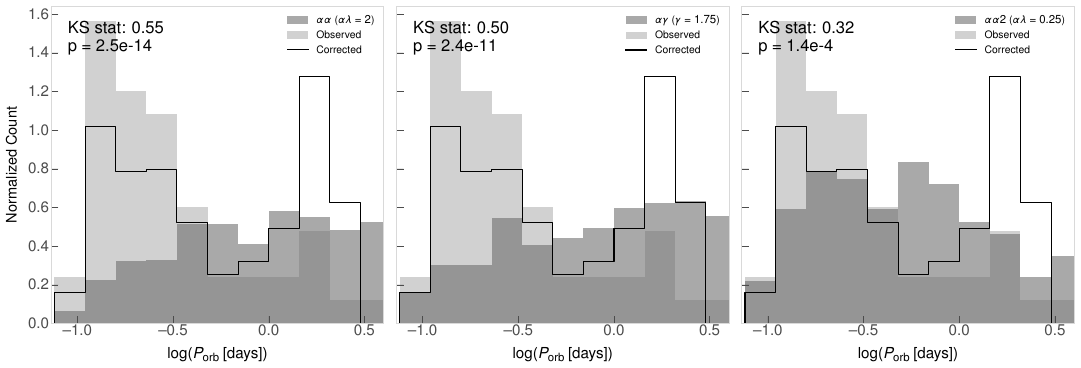}
    \caption{Comparison between our completeness-corrected observed PCEB period distribution (black curve; classified CVs removed) and BPS model predictions (gray histograms) filtered using observational cuts matching the initial sample. The theoretical models span multiple CE prescriptions and initial binary population assumptions.}
    \label{fig:BPS_comparison}
\end{figure*}

The theoretical expectation for PCEB orbital periods is provided by BPS models. In particular, we consider the synthetic WDMS binary sample from \citet{Toonen_2013}, hereafter \citetalias{Toonen_2013}. This synthetic sample excludes any systems that have reached contact and turned into CVs, and so, the BPS distributions we compare to contain only detached post-CE WDMS binaries. We omit any systems that are classified as CV, while keeping in mind that there is a high probability that this subsample is contaminated by unclassified CVs.

 \citetalias{Toonen_2013} explored a range of CE efficiencies and found that the period distribution of PCEBs is highly sensitive to $\alpha\lambda$. Here, $\lambda$ is the envelope-structure (binding energy) parameter that relates the donor's envelope binding energy to its mass and radius \citep{deKool_1990}. High-efficiency models ($\alpha\lambda \approx 1-2$) produce a broad distribution extending to long orbital periods (tens of days), while low-efficiency models ($\alpha\lambda \approx 0.2-0.3$) produce a much more compact period distribution concentrated at shorter periods. They conclude that selection effects do not significantly shape the period distribution of visible PCEBs, and that the dearth of long-period systems is mostly explained by a low $\alpha\lambda$ efficiency of the main evolutionary channel. Their illustrative best-fit model ($\alpha\lambda=0.25$) thus demonstrates the need for efficiencies in the range $\alpha\lambda\sim0.1–0.3$. In that scenario, the vast majority of PCEBs emerge with $P_{\rm orb}\lesssim10\,$d, and the distribution peaks sharply at $P_{\rm orb}\approx0.3-0.5$\,d ($\sim7-12$\,h).

To place our empirical PCEB period distribution in context, we compare it with predictions from BPS models. We focus on simulations from \citetalias{Toonen_2013} discussed above. Their simulations incorporate detailed prescriptions for binary interaction physics, Galactic star formation history, and population synthesis, making them a robust framework for interpreting the observed distribution.

To ensure a meaningful comparison, we applied the same detection filters to the simulated data as in our observed sample. Specifically, we impose color selection based on Gaia photometry, a contrast-ratio cut, and a mass-ratio constraint for our observational biases. These cuts mirror the selection criteria in the light curve analysis of our sample.

The BPS models differ in their treatment of CE evolution. The $\alpha\alpha$ and $\alpha\alpha2$ models adopt the energy-based CE ejection formalism, in which the envelope is ejected at the expense of orbital energy. They differ only in their CE ejection efficiency, with $\alpha\lambda = 2.0$ (efficient) and $\alpha\lambda = 0.25$ (inefficient), respectively. In contrast, the $\alpha\gamma$ model employs angular momentum conservation (the $\gamma$-formalism, with $\gamma = 1.75$) to govern envelope ejection, except when CE is triggered by a tidal instability, in which case it reverts to the energy prescription.

Figure~\ref{fig:BPS_comparison} compares our completeness-corrected PCEB period distribution (black step; classified CVs removed) with the predicted distributions from three BPS models ($\alpha\alpha$, $\alpha\gamma$, and $\alpha\alpha2$, gray histograms), each filtered using observational cuts matching the initial sample. The observed uncorrected distribution is shown as reference (light gray). None of the models reproduces the data within statistical uncertainties, KS statistics indicate significant tension in all cases, although the inefficient model ($\alpha\lambda = 0.25$) provides the closest match.

The drop on the short-period side of the corrected distribution is a result of our removal of systems that are classified as CVs. As soon as Roche-lobe overflow begins, the binary is excluded from the PCEB sample, removing most of the systems with periods below $\sim$2\,h. While the corrected distribution is bi-modal, generating a ``gap'' in the synthetic population would require either two distinct formation channels (which our mass‐ratio distribution does not support) or two separate CE efficiencies, raising the question of what physics might drive such a bifurcation.

The apparent bimodality in our observed orbital period distribution may reflect different evolutionary phases within the PCE population. The longer-period systems likely represent detached PCEBs that have not yet initiated mass transfer. In contrast, the shorter-period peak predominantly consists of systems that have already evolved into CVs, reflecting a further evolutionary stage characterized by ongoing mass transfer. Since approximately half of our short-period systems are indeed classified as CVs in the literature (see Figure~\ref{fig:ellipsoidals_period_distribution} below), one might suspect that many of the remaining short-period systems used in this comparison could be yet unclassified CVs. Further classification is required to reconcile the observed distribution with theoretical predictions, as the two peaks likely represent different evolutionary states rather than variations in CE efficiency alone.

\subsection{Sinusoidal Variables}\label{subsec:ellipsoidal modulations}

\begin{figure}
    \centering
    \includegraphics[width=1\linewidth]{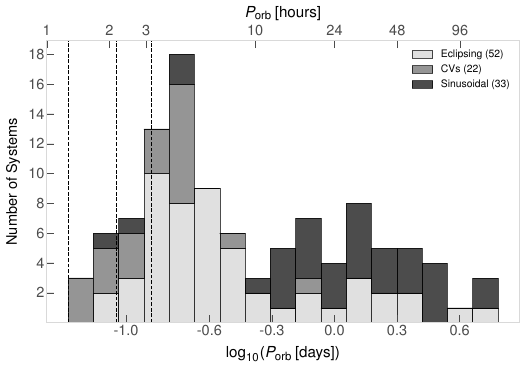}
    \caption{Stacked histogram of the orbital period distribution (in log hours) for three sub-classes:
    eclipsing PCEBs (light gray, N = 52), CVs (medium gray, N = 22),  and sinusoidal variables (dark gray, N = 33). Vertical dashed lines mark the minimum observed CV period ($\approx$  80\,min ) and the canonical lower ($\sim2$\,h) and upper ($\sim3$\,h) boundaries of the CV ``period gap''.}
    \label{fig:ellipsoidals_period_distribution}
\end{figure}

In addition to the 74 eclipsing systems, our periodicity search identified 33 WDMS binaries showing strong sinusoidal variability, likely caused by effects such as ellipsoidal modulation or reflection. These systems show characteristic sinusoidal light curves with orbital periods ranging from $\approx1$ to $\approx65$~hours, but lack the clear eclipse features that define our primary sample (primary, secondary, box or v-shaped signals). We use ``sinusoidal'' as a phenomenological label and do not classify the mechanism. While some of these modulations are consistent with ellipsoidal distortion or reflection, they can also arise from accretion-related processes (\textit{e.g.,} superhumps or bright-spot modulations), or starspots. The measured photometric period may trace the orbital period or its first harmonic (and in superhumping systems can be slightly offset). We therefore report the measured periods without correction and avoid further classification based on TESS data alone. For literature-classified CVs, we adopt their confirmed orbital periods when available. 

Figure~\ref{fig:ellipsoidals_period_distribution} shows the period distribution of all periodic systems in our sample, subdivided into three categories, confirmed CVs (22), eclipsing PCEBs (52), and sinusoidal variables (33). The sinusoidal systems are distributed across the entire period range, from the shortest periods overlapping with known CVs to the longest periods in our survey. Because ellipsoidal signals are symmetric across the orbit, the detected periods may correspond to a harmonic of the true orbital period. We report the measured periods without correction, except for literature-classified CVs, where we adopt the confirmed orbital periods when available.

Ellipsoidal variability occurs when the secondary star is tidally distorted, causing the observed flux to vary with changes in its projected area and surface brightness over the orbit. Additional reflection or emission from the heated hemisphere further modulates the light curve. Such signatures are expected in compact WDMS systems with large Roche-lobe fill factors, and they persist even when eclipses are undetectable at the TESS cadence or precision.

Few effects can suppress resolvable eclipses while preserving strong ellipsoidal/reflection signals. Orbital inclinations just below the eclipse threshold, grazing or shallow events, photometric dilution in crowded pixels that hides shallow eclipses while leaving the broader sinusoid detectable, and cadence smearing at 10 to 30\,min sampling that reduces the apparent depth and duration of brief events.

Importantly, 9 of the known CVs appear as sinusoidal variables rather than clear eclipsers in the TESS data, demonstrating that this variability class can include systems with confirmed ongoing mass transfer. It is therefore plausible that a fraction of the sinusoidal variables are interacting or near-interacting binaries (\textit{e.g.,} pre-CVs or weakly accreting systems) whose eclipses are either grazing, diluted, or cadence-smeared below our $\mathrm{SNR}>5$ eclipse threshold.

 This possibility is supported by previous studies (e.g., \citealt{Drake_2014a, Catalina_PCEBs}) where systems showing strong ellipsoidal variability were later revealed to host shallow eclipses when observed with higher photometric precision. In our case, the TESS pixel scale may simply dilute or smooth out shallow eclipses, specifically for partially grazing configurations or in systems with smaller WDs.

These sinusoidal variables represent an important subclass of post-CE systems. While they do not formally meet our eclipse-based detection criteria, they nonetheless exhibit strong evidence for close binary interactions and compact orbital separations. Including them in the broader PCEB population provides a more complete view of the evolutionary outcomes of CE evolution. A dedicated follow-up campaign in the blue or near-ultraviolet, with higher temporal resolution and sensitivity, would likely confirm the eclipsing nature of many such systems. Additional photometric observation with a ground-based observatory, such as the Large Array Survey Telescope \citep[LAST;][]{Ben_Ami_2023,Ofek_2023}, could easily uncover eclipses diluted in the TESS pixels to increase the statistics of the eclipsing sample.

\subsection{The PCEB Fraction Among WDMS Binaries}
\label{sec:pceb_fraction}

We searched TESS light curves for $N_{\rm tot}=583$ WDMS binaries and identified two observational classes:
(i) \emph{short-period} systems with $P_{\rm orb}<13$\,h (interpreted as interacting PCEBs), and
(ii) \emph{long-period} systems with $P_{\rm orb}\ge 13$\,h (up to $P_{\rm orb} \approx 6$\,d; non-interacting PCEBs).
We detected $N_{\rm obs,<13}=60$ systems in the short-period class and $N_{\rm obs,>13}=14$ in the long-period class.

The observed fractions are therefore
\[
f_{\rm obs,<13}=0.103\pm0.013,
\quad
f_{\rm obs,>13}=0.024\pm0.0063,
\]
where the uncertainties are $1\sigma$ binomial errors.

To infer the \emph{intrinsic} fractions we correct for the geometrical selection and detectability. We compute an \emph{interval-averaged detection probability} directly from our bias-corrected period distribution:
\[
\overline{p}_{\rm det}(S) \;\equiv\;
\frac{\sum_{b\in S} D_b}{\sum_{b\in S} D_b^{\rm corr}},
\]
where $D_b$ is the observed count in bin $b$, $D_b^{\rm corr}$ is the bias–corrected count in the same bin, and $S$ is the set of bins in the period interval of interest ($P<13$\,h or $P\ge 13$\,h). This ratio is the appropriate average of the detectability function over the interval.

With $\overline{p}_{\rm det,<13}=0.78$ and $\overline{p}_{\rm det,>13}=0.15$, the intrinsic fractions follow from
\[
f_{\rm true}(S)\;=\;\frac{f_{\rm obs}(S)}{\overline{p}_{\rm det}(S)}.
\]
Resulting in,
\[
f_{\rm true,<13} = 0.132 \pm 0.0166,
\qquad
f_{\rm true,>13} = 0.17 \pm 0.042,
\]
where the quoted uncertainties propagate only the binomial errors on $f_{\rm obs}$ (i.e., $\sigma_{f_{\rm true}}\simeq \sigma_{f_{\rm obs}}/\overline{p}_{\rm det}$). The total intrinsic PCEB fraction in our WDMS sample is then
\[
f_{\rm true,tot} \;=\; f_{\rm true,<13}+f_{\rm true,>13}
\;=\; 0.298 \pm 0.045.
\]

Finally, we note that spatially resolved WDMS pairs are not part of our target list and are therefore excluded from these fractions by construction.

For spatially unresolved WDMS, we find $f_{\rm PCEB}= 29.8 \pm 4.5\%$
, which is consistent with the SDSS estimate of 
27$\pm2\%$ \citep{SDSS_PCEBs}.  

\subsection{Comparison with UV-calibrated CV accretion rates}

\begin{figure*}
  \centering
  \includegraphics[width=\textwidth]{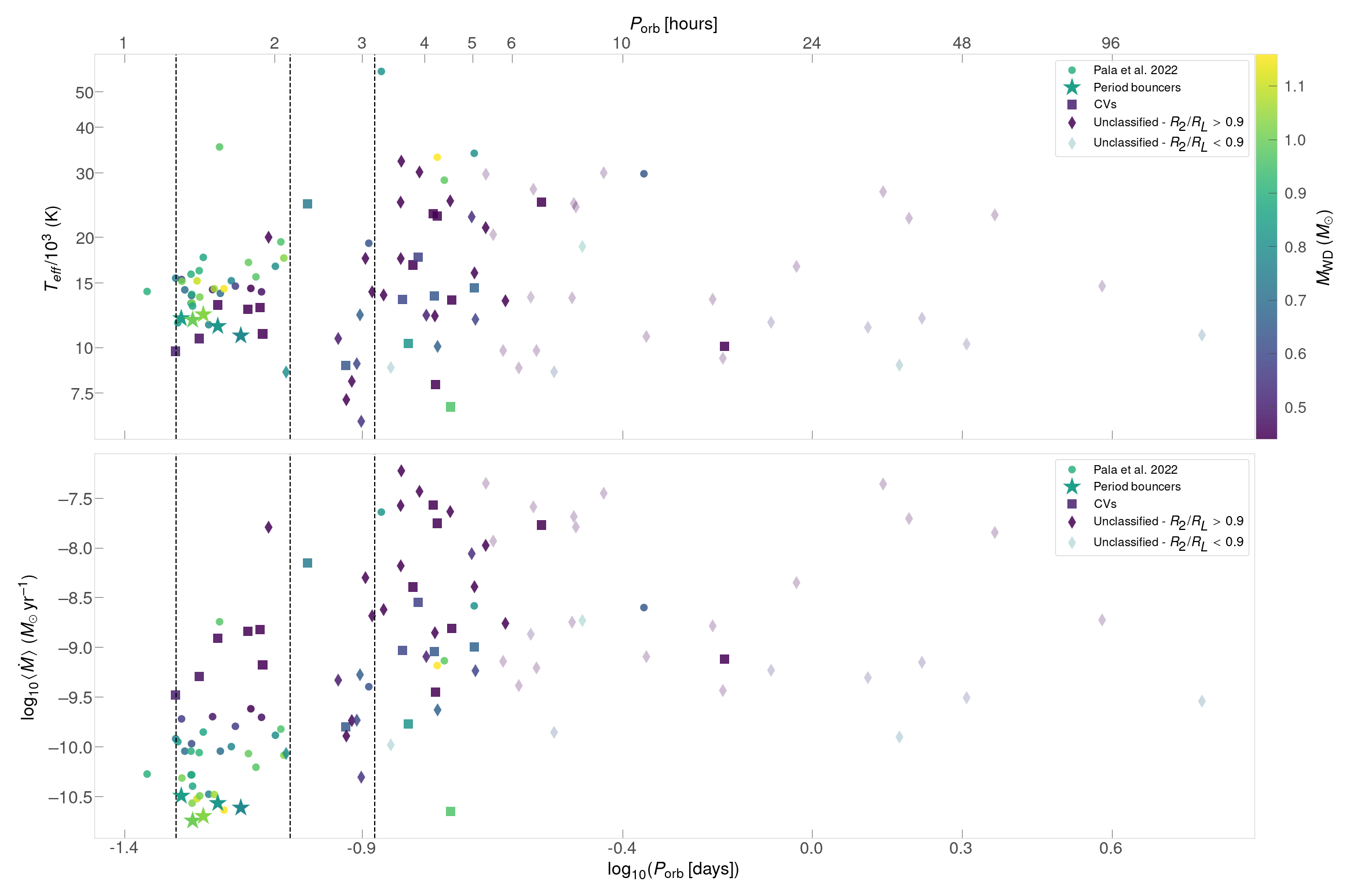} 
  \caption{Comparison of our eclipsing WDMS systems (52 unclassified PCEBs + 22 CVs; sinusoidal variables excluded) to the UV-calibrated CV locus from the HST+Gaia sample of \citet{Pala_22}. \textit{Top}: WD effective temperature versus orbital period. \textit{Bottom}: equivalent long-term accretion rate $\langle\dot M\rangle$ (inferred from the WD's $T_{\rm eff}$ following the UV calibration) versus orbital period. Vertical dashed lines mark the minimum observed CV period ($\approx$  80\,min ) and the canonical lower ($\sim2$\,h) and upper ($\sim3$\,h) boundaries of the CV ``period gap''. Unclassified eclipsers at $P_\text{orb} \approx 3  -  6$\,h lie on the same $T_{\rm eff} \ - \ P_{\rm orb}$ sequence as CVs above the period gap, indicating that many systems are near, or already interacting, even without a CV classification.}
  \label{fig:temp_period}
\end{figure*}

Figure~\ref{fig:temp_period} compares our eclipsing WDMS sample (52 non-CV PCEBs + 22 CVs; sinusoidal variables excluded) to the HST+Gaia CV locus of \citet{Pala_22}. Our non-CV eclipsers at orbital periods $\simeq3-6$\,h lie on the same $T_{\rm eff}$--$P_{\rm orb}$ and $\langle\dot M\rangle$--$P_{\rm orb}$ sequences traced by the UV-calibrated CVs above the gap, \textit{i.e.}, $T_{\rm eff}\sim15-30$\,kK and $\log_{10}\langle\dot M\rangle\sim-9.5$ to $-8$. Suggesting that many of these systems are near- or already-interacting despite lacking a CV classification. Our CV subsample reproduces the canonical gap behavior of hotter, higher-$\langle\dot M\rangle$ systems above 2 -- 3\,h and cooler, lower-$\langle\dot M\rangle$ systems below it. We deliberately keep $\langle\dot M\rangle$ for unclassified-CV eclipsers as an \emph{upper-limit, ``equivalent'' rate}, because the $T_{\rm eff}\!\to\!\langle\dot M\rangle$ relation assumes long-term accretional heating. Detached systems may be hot for other reasons. 

While the orbital periods are precise, individual points in Figure~\ref{fig:temp_period} carry large uncertainties. WD mass typical error is $\approx 0.3\, M_\odot$, and since $L_{\rm acc}\!\propto\!GM_{\rm WD}\langle\dot M\rangle/R_{\rm WD}$, these mass errors propagate directly into $\langle\dot M\rangle$. Likewise, uncertainties in the WD cooling age ($\approx 0.1\,\rm{Gyr}$) and atmosphere model assumptions affect the mapping between $T_{\rm eff}$ and the long-term accretion heating. Therefore, our comparison is appropriate when considering \emph{population-level trends} rather than individual objects. The ensemble patterns are robust, but single-point values of $\langle\dot M\rangle$ and $T_{\rm eff}$ should be interpreted with caution.

These patterns mirror the UV-calibrated relations established by \citet{Pala_22}. Hotter WDs and higher $\langle\dot M\rangle$ lie above the gap. Cooler, lower-$\langle\dot M\rangle$ systems appear below it, and a distinct low-$\langle\dot M\rangle$ period-bouncer branch is located near the minimum. One main difference is coverage above the gap, whereas the UV-calibrated sample contains relatively few systems in this regime, the majority of our binaries inhabit it, providing improved leverage on the above-gap population.

Our WD mass distribution is skewed to low masses ($\sim$ 0.3 - 0.5\,$M_\odot$; Figure~\ref{fig:mass_distribution}), which likely reflects our Gaia-XP selection (Section~\ref{subSec:selection_effects}). For a fixed long-term $\langle\dot M\rangle$, the lower-mass WD accretion features are intrinsically weaker. This might explain why many of our near-contact eclipsers at $P \sim$ 3 - 6\,h sit on the CV $T_{\rm eff}$ -- $P$ and $\dot{M}$ -- $P$ sequences in Figure~\ref{fig:temp_period}, yet lack firm CV classifications.

Empirical consequential angular momentum loss (eCAML; \citet{Schreiber_2015}) models predict the secular average mass-transfer rate as a function of orbital period, $\langle \dot M \rangle (P_{\rm orb})$, and the resulting accretion-heated WD temperature. Our systems are cooler than the eCAML expectations at a given orbital period, yet when mapped to $\langle \dot M \rangle$ across $P_{\rm orb}\sim$ 3 - 6\,h, our eclipsers span the $\langle \dot M \rangle$ range outlined by the eCAML models in \cite{Pala_22}. This alignment in $\langle \dot M \rangle$ alongside cooler $T_{\rm eff}$ is consistent with our skewed mass distribution, which reduces accretion heating at a fixed $\langle \dot M \rangle$.

Beyond the cataloged CVs, our sample reveals a majority of compact WDMS systems that already lie on the CV locus but remain unclassified, potentially due to a lack of detectable outbursts or unambiguous accretion signatures. Only 22 out of 107 systems are CV-classified, whereas $\sim 80\%$ are eclipsing or sinusoidally modulated WDMS without a CV label. Within the eclipsing set alone, $\sim 70\%$ are not classified as CVs. Given the inclination and detectability biases, this imbalance likely understates the intrinsic prevalence of weakly accreting or pre-CV systems. Together with the corrected period distribution, whose short-period peak is dominated by near-contact binaries, we conclude that there is a large reservoir of pre-CV or weakly interacting PCEBs that feed the CV population but are missed by eclipse-only or outburst-selected searches.

\section{Conclusions}\label{sec:conclusions}
We cross-matched the new Gaia DR3 WDMS catalog with TESS photometry and analyzed the orbital period distribution of this population. Our search identified a total of 107 periodic systems, including 74 eclipsing binaries and 33 presenting only sinusoidal modulation, of which 32 eclipsing systems are newly reported. The observed distribution is strongly peaked at $\approx 4$\,h with a relatively narrow width ($\approx 1.8$\,h), and an additional statistically significant long-period tail extending up to a few days. After applying vetting and completeness corrections, we derived the intrinsic orbital period distribution for this population. A two-component model provides a good description: a log-normal peak and a rising linear tail that begins near $\approx13$\,h. This implies that while the majority of systems emerge from the CE phase with short orbital periods or evolve further toward shorter orbits, a non-negligible fraction is formed at wider separations of up to a few days. The short-period peak may also be contaminated by unclassified, weakly accreting pre-CVs. We estimate that roughly one-third of unresolved WDMS systems within our catalog ($29.8\%\pm4.5\%$) are close, post-CE binaries, a fraction consistent with earlier SDSS-based studies ($\sim27\%$; \cite{SDSS_PCEBs}). Importantly, our sample includes 22 previously known CVs and a number of systems that exhibit nearly Roche-lobe filling configurations. The peak feature of the distribution remains even when CVs are omitted, confirming that we are observing the immediate precursors and early members of the CV population.

As shown in Figure~\ref{fig:BPS_comparison}, none of the tested BPS prescriptions reproduces both the sharp short-period peak and the extended long-period tail simultaneously: the $\alpha\alpha$ and $\alpha\gamma$ models predict broader, long-period-heavy distributions than observed, whereas $\alpha\alpha\,2$ concentrates systems more strongly at short periods but underproduces the tail. However, the short-period peak remains heavily populated by systems with high Roche-lobe filling factors. Since interacting systems are removed from the BPS samples, this indicates that a substantial fraction of the peak may still be composed of pre-CVs or weakly accreting systems. 

Two interpretations, therefore, remain viable: either CE efficiency must vary in a way that naturally produces both a compact, interacting-rich peak and a detached long-period tail, or the peak is more strongly contaminated than accounted for. In either case, our results highlight that current CE prescriptions do not yet capture the full complexity of the observed period distribution.
Further study of detached WDMS population is needed, over a wide range of orbital separations, is required to address this ambiguity \citep[e.g.,][]{Hallakoun2024, Rekhi2024, Rekhi2025, Ironi2025, Shahaf2025}.

Future progress will depend both on clarifying the composition of the short-period peak and on expanding the sample across the full period range. Targeted spectroscopy and time-series photometry of high–fill–factor systems are necessary to distinguish detached PCEBs from pre-CVs or weak accretors, thereby refining the detached-only distribution for robust CE-model comparison. Most of the WDMS binaries in the Gaia catalog lack TESS observations, and systematic efforts to recover sinusoidal variable systems and spectroscopic binaries will provide a less inclination-biased view of the distribution. Together, these steps will enable us to test whether modifications to CE efficiency are required to generate both the compact, interaction-dominated peak and the detached long-period tail, or whether stricter removal of pre-CVs from the peak resolves the discrepancy.

\begin{acknowledgments}
NH acknowledges support from the Planning \& Budgeting Committee of the Israeli Council for Higher Education.
SS, JL, and HWR acknowledge support from the European
Research Council for the ERC Advanced Grant [101054731]. SS was also supported by a Benoziyo Prize postdoctoral fellowship.
ST acknowledges support from the Netherlands Research Council NWO (grant VIDI 203.061).

This work has made use of data from the European Space Agency (ESA) mission Gaia (\url{https://www.cosmos.esa.int/gaia}), processed by the Gaia Data Processing and Analysis Consortium (DPAC; \url{https://www.cosmos.esa.int/web/gaia/dpac/consortium}). Funding for the DPAC has been provided by national institutions, in particular the institutions participating in the Gaia Multilateral Agreement.

This paper includes data collected with the TESS mission, obtained from the MAST data archive at the Space Telescope Science Institute (STScI). Funding for the TESS mission is provided by the NASA Explorer Program. STScI is operated by the Association of Universities for Research in Astronomy, Inc., under NASA contract NAS 5–26555.

This research has made use of the SIMBAD database, operated at CDS, Strasbourg, France \citep{Wenger_2000_Simbad}.

\end{acknowledgments}

\begin{contribution}



YS led and carried out the analysis and prepared the manuscript. NH, SBA, and SS contributed to the study’s conceptualization and validation. JL provided the WDMS catalog, including the mass and temperature estimates. HWR contributed to the discussion of the selection function. ST supplied the BPS sample specifically tailored for this work and contributed to the discussion of the theoretical models. All authors contributed to the discussion and provided comments on the manuscript.

\end{contribution}

%
\facilities{Gaia, TESS}


\software{Astropy \citep{astropy:2013,astropy:2018,astropy:2022}, fBLS \citep{Shahaf_2022},
          Lightkurve \citep{lightkurve:ascl},
          PHOEBE \citep{Prsa2016}, WD\_models (\url{https://github.com/SihaoCheng/WD_models})}


\appendix
\begin{table*}[t]
\centering
\caption{Key properties for the TESS subset (the full table will be available online).}
\label{tab:sample}
\setlength{\tabcolsep}{3pt}
\renewcommand{\arraystretch}{0.9}
\scriptsize
\begin{tabular}{llrrrrrrrrrr}
\hline\hline
Gaia DR3 source\_id & TIC & RA [deg] & Dec [deg] & Tmag [mag] & G [mag] & BP$-$RP [mag] & $T_{\rm eff}$ [K] & $T_{\rm cool}$ [K] & $M_{\rm WD}$ [$M_\odot$] & $\sigma_{M_{\rm WD}}$ [$M_\odot$] & $M_2$ [$M_\odot$] \\
\hline
5164304518213598720 & 38399026  &  54.054361 &  -9.566 293 & 15.44 & 16.80 & 1.46 &  9904 & 12588 & 0.807 & 0.430 & 0.425 \\
4788741548375134336 & 219244444 &  65.273180 & -48.651962 & 12.63 & 13.64 & 1.87 &  7431 &  8569 & 0.622 & 0.213 & 0.216 \\
4795131772517207552 & 151580578 &  87.955811 & -47.145865 & 14.90 & 15.97 & 2.05 &  8763 &  9842 & 0.478 & 0.161 & 0.337 \\
4795838002579632128 & 235072034 &  86.096490 & -47.171550 & 14.83 & 15.41 & 1.18 & 12546 & 27911 & 0.820 & 0.120 & 0.339 \\
4756545648896550400 & 149931175 &  89.476259 & -63.664190 & 15.87 & 16.93 & 2.02 &  9763 & 14726 & 0.815 & 0.165 & 0.343 \\
6604630592725421440 & 2056016660& 345.554651 & -31.933080 & 16.05 & 16.21 & 0.47 & 20037 & 92493 & 0.905 & 0.327 & 0.404 \\
6604911414866797056 & 229065739 & 344.959349 & -31.144723 & 16.00 & 16.94 & 1.75 & 14102 & 16734 & 0.412 & 0.327 & 0.540 \\
6606222964734715264 & 12969661  & 345.880371 & -29.820095 & 15.30 & 16.51 & 1.67 & 12278 & 10372 & 0.770 & 0.419 & 0.415 \\
6616573488025635456 & 80083663  & 331.144424 & -30.018363 & 13.90 & 14.88 & 1.77 & 10178 & 13104 & 0.571 & 0.083 & 0.330 \\
5337884487666708096 & 308282854 & 167.442864 & -59.163964 & 15.27 & 16.18 & 1.89 & 10566 & 12009 & 0.526 & 0.376 & 0.369 \\
\hline
\end{tabular}
\end{table*}

\section{Appendix information}

\subsection{Comparison to previous works}\label{sec:comparison_to_previous_work}

\begin{figure}
    \centering
    \includegraphics[width=1\columnwidth]{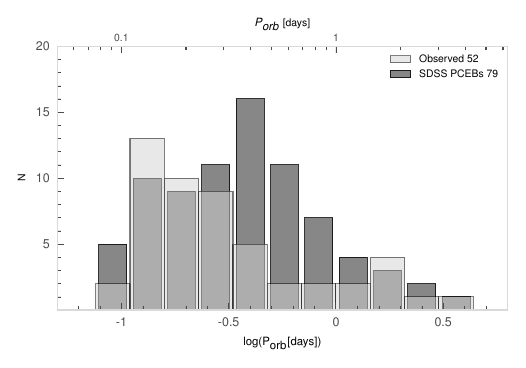}
    \caption{Comparison of the eclipsing systems from our sample with the SDSS PCEB sample. KS Statistic = 0.2609 with a $p$-value $\approx$ 0.002 for a sample of 100 systems.}
    \label{fig:SDSS_comparison}
\end{figure}

Our empirically derived period distribution can be compared to that of the SDSS PCEB sample of \citet{SDSS_PCEBs}, who identified a large, homogeneously selected population of WDMS binaries, and derived a bias-corrected period distribution spanning 1.9\,h to 4.3\,d. They reported a log-normal shape peaking at $P_{\mathrm{orb}} \approx 10.3$\,h with a width of $\sigma_{\log P}\approx 0.41$\,dex ($\approx 2.6\,$h). In contrast, our log-normal component peaks at a shorter period, $P_{\mathrm{orb}} \approx 4$\,h, and is significantly narrower, with $\sigma_{\log P}\approx 0.26$\,dex ($\approx 1.8\,$hr). Figure~\ref{fig:SDSS_comparison} compares the orbital-period distribution of our eclipsing sample to the SDSS RV-selected PCEB sample, showing a clear shift toward shorter periods. This discrepancy likely arises from differences in the sample selection and the detection sensitivity. The SDSS survey identified PCEBs via RV variations, which are more sensitive to longer period systems compared to the eclipse method. Our sample includes only eclipsing binaries identified in TESS photometry, which are strongly biased toward shorter periods due to the geometric eclipse probability. Additionally, our sample is confined to the \citet{Li2024} catalog and its cuts, which may over-represent lower-mass, short-period systems.  While we apply a detailed completeness correction to mitigate this bias, the peak at shorter periods remains. Despite these differences, both surveys agree on the key feature, the majority of PCEBs have orbital periods well below one day. Such a discrepancy might break when excluding the interacting PCEBs, which requires CV classification.

\subsubsection{Catalina}
The Catalina Real-Time Transient Survey provided long-baseline optical
light curves for thousands of WDMS binaries. \cite{Catalina_PCEBs}
searched 835 spectroscopically confirmed SDSS WDMS systems and found
29 eclipsing PCEBs; 17 were already known and 12 were new. They noted
that the orbital period distribution of the eclipsing PCEBs is skewed
toward short periods compared with the overall SDSS PCEB distribution
because the probability of observing an eclipse decreases with orbital
separation. They detected two eclipsing systems with periods longer than 1.9\,d,
extending the known range of eclipsing PCEBs. A later
study using Catalina data discovered 14 additional eclipsing WDMS systems
and numerous ellipsoidally modulated, non‑eclipsing PCEBs \citep{Parsons_2015}. The periods of
the ellipsoidal systems cluster tightly around 4\,h because their M2–M3
donor stars are close to filling their Roche lobes. Our distribution is broadly consistent with the
Catalina eclipsing PCEB sample, both exhibit a strong peak at a few hours
and a steep decline toward longer periods. In contrast to Catalina ellipsoidal modulated binaries, we find sinusoidal modulations (see Section~\ref{subsec:ellipsoidal modulations}) across the entire distribution. Figure~\ref{fig:ellipsoidals_period_distribution} introduces all the systems with a periodic signal, divided into 3 sub-classes: CVs (light gray), eclipsing PCEBs (medium gray), and non-eclipsing systems with sinusoidal modulations.  

\subsubsection{ZTF}
\cite{Keller_2021} cross‑matched Gaia WD
candidates with ZTF DR3 light curves, and performed a BLS
search for periodic dimming events. They discovered 18 new WD
binaries, 17 of which are eclipsing. Table~1 of that work lists the orbital
periods and classifications of the systems. The detected periods span
1.03 -- 5.72\,h, and the majority of the WDMS systems have periods between 1.4 -- 3.6\,h. The authors noted that their search is most sensitive to short periods due to the geometrical constraint and that the cadence of ZTF limits
detections of longer eclipses. Several systems with periods between  2.4 -- 5.72\,h harbor extremely low‑mass WDs, while the shortest periods include CVs and AM~CVn systems. 

A complementary ZTF study by \cite{LiAndZhang_2024} analyzed the DR19
light curves of  around 200 WDMS binaries exhibiting Balmer emission lines in
SDSS or LAMOST spectra. They identified 55 PCEBs with periods
ranging from 2.2\,h to 81.6\,h (0.09 -- 3.38\,d) and noted that only six
were previously unknown. Three of the newly discovered
systems are eclipsing. Because this sample is selected via chromospheric
emission rather than eclipses, it includes more intermediate‑period
systems compared with our sample.
In summary, the ZTF results reinforce the dominance of short-period systems, consistent with the peak we observe around a few hours. However, their strong selection bias toward short periods prevents them from probing the extended long-period tail that emerges in our TESS-based distribution. The presence of extremely low-mass WDs and AM~CVn systems in the ZTF sample further supports the view that surveys sensitive to such systems, including ours, naturally recover shorter-period PCEBs compared to the RV-selected SDSS sample.

\subsection{Potential Post–Common Envelope Binaries Catalog}

\citet{Kruckow_2021} compiled a catalog of 839 candidate PCEBs by
assembling data from a wide range of observational campaigns. They
inspected roughly a thousand published studies on systems suspected of
CE evolution and merged those objects into a unified dataset. The catalog includes systems discovered via detailed spectroscopy, eclipsing light curves, nova outbursts and
other methods. To be included, a binary had to contain at least one hydrogen‑depleted component (a WD or a hot subdwarf (sd)) and have fundamental parameters published in the literature. In practice, this meant requiring an orbital period shorter than 100~days, and at least having limits on the masses of both components. The authors divided the sample into three major classes: double WDs (DWD), WDs with non‑degenerate companions, and sdB/O binaries; and explicitly excluded systems where the CE donor is a neutron star or a black hole, because supernova explosions erase clear CE signatures.

\begin{figure}
    \centering
    \includegraphics[width=1\columnwidth]{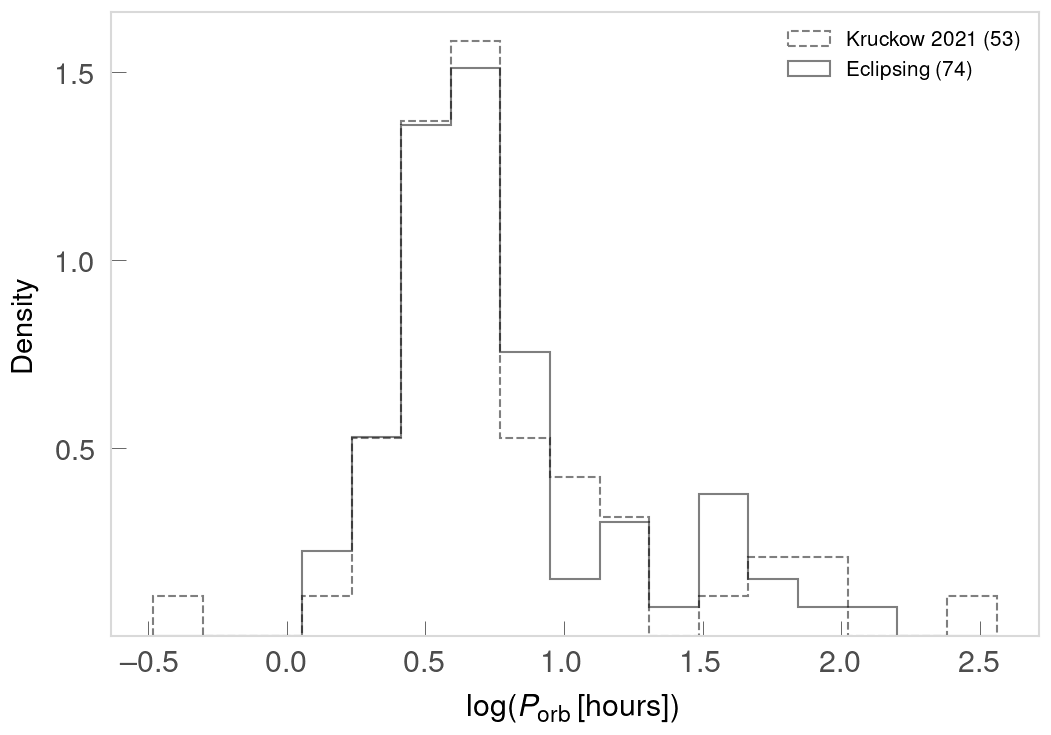}
    \caption{Orbital period distribution comparison of our eclipsing sample (74 systems, solid line) and the subset of potential PCEBs from the \citet{Kruckow_2021} catalog (53 systems, dashed line) selected to have MS companions with masses below $0.8$\,\msun. Both distributions exhibit a peak around a few hours, highlighting the typical compact nature of PCEBs.}
    \label{fig:kurkuw}
\end{figure}

Selecting systems with MS companions less massive than $0.8$\,\msun\ yields a subset of 53 objects suitable for comparison with our eclipsing sample, demonstrated in Figure~\ref{fig:kurkuw}. Both distributions are consistent. Only eight systems overlap
between our samples.


\subsection{Roche geometry classification}
\label{sec:roche_method}
For each system, we evaluate the secondary's Roche-lobe radius,
\begin{equation}
    \frac{R_{L,2}}{a} = \frac{0.49 q ^{\frac{2}{3}}}{0.6q ^{\frac{2}{3}} + ln(1+q ^{\frac{1}{3}})}
\end{equation}
 \citep{Eggelton_1983}. 
The orbital separation $a$ follows from Kepler’s law at the observed period.
The stellar radius $R_2$ is estimated from a mass-radius relation for M dwarfs 
($R_2\simeq M_2^{0.8}R_\odot$; \citet{1990_Jean_Pierre}).
We define the fill factor $f_{\rm fill}=R_2/R_{L,2}$ and classify systems as
near-contact ($f_{\rm fill}\ge0.9$), or detached ($f_{\rm fill}<0.9$).
Results are insensitive to the exact threshold: adopting $f_{\rm fill}=0.85$ or $0.95$ does not change the qualitative separation or our conclusions.
\clearpage
\startlongtable
\begin{deluxetable}{cccccccc}
\tablecaption{Eclipsing and classified systems within our sample.\label{tab:detected_systems}}

\tablewidth{0pt}
\tabletypesize{\scriptsize} 
\tablehead{
\colhead{Gaia DR3 ID} & \colhead{RA} & \colhead{Dec} & \colhead{G} & \colhead{$G_\text{BP} - G_\text{RP}$} & \colhead{Period}  & \colhead{Classification} \\
\colhead{} & \colhead{(deg)} & \colhead{(deg)} & \colhead{(mag)} & \colhead{(mag)} & \colhead{(d)} & \colhead{} & \colhead{}
}
\startdata
1223491257550872192 & 238.55136 & 27.36463 & 17.154 & 1.072 & 0.052709 &   EBCV \\
1289860214647954816 & 225.67042 & 33.57317 & 17.261 & 0.338 & 0.058908 &   CV \\
6896767366186700416 & 318.03874 & -8.82707 & 16.816 & 0.553 & 0.064153 &   CV \\
426306363477869696 & 17.55491 & 60.07646 & 16.313 & 0.648 & 0.073646 &   CV \\
5524430207364715520 & 134.63786 & -41.79793 & 16.639 & 0.781 & 0.078058 &   CV \\
4697621824327141248 & 25.25221 & -67.89088 & 17.213 & 0.781 & 0.078912 &   CV \\
3195473019892503552 & 62.57005 & -8.57213 & 17.428 & 0.196 & 0.081105 &   EB \\
4588070505828585984 & 276.99807 & 29.09155 & 16.524 & 2.043 & 0.08796 &   Unclassified \\
6684490840966796928 & 294.64945 & -46.21598 & 16.807 & 1.036 & 0.097193 &   CV \\
6468246487517124736 & 308.40486 & -56.56259 & 16.63 & 1.236 & 0.101549 &   CV \\
4701214616008598144 & 34.80301 & -63.11582 & 15.981 & 0.841 & 0.111917 &   Unclassified \\
755705822218381184 & 156.61466 & 38.75055 & 16.339 & 2.146 & 0.115924 &   CV \\
1635581274974672768 & 256.30386 & 66.12709 & 15.452 & 2.201 & 0.116195 &   Unclassified \\
1685720070351584000 & 197.40734 & 69.55443 & 15.904 & 2.191 & 0.119177 &   Unclassified \\
3662988495054111232 & 204.06557 & 0.29175 & 16.695 & 1.523 & 0.121969 &   EB \\
2287248210301342336 & 339.31498 & 82.17435 & 15.4 & 1.745 & 0.123783 &   Unclassified \\
4028597414327329664 & 182.54089 & 33.78879 & 15.507 & 2.348 & 0.124501 &   EB \\
5433756094759976192 & 149.50562 & -37.57242 & 16.344 & 0.892 & 0.126933 &   EB \\
5362212384974013184 & 165.82197 & -48.58444 & 16.27 & 1.225 & 0.130923 &   EB \\
2856883739877737216 & 4.46252 & 27.85933 & 17.271 & 0.492 & 0.138056 &   EB \\
2273583445431091584 & 316.64474 & 72.52089 & 16.042 & 1.684 & 0.142721 &   Unclassified \\
2025873096433233664 & 290.05891 & 27.37168 & 15.559 & 0.274 & 0.149363 &   EB \\
580790014913812608 & 137.05016 & 6.07252 & 17.069 & 0.545 & 0.149456 &   EB \\
4911920901187418112 & 29.47929 & -54.51009 & 16.992 & 1.175 & 0.149823 &   Unclassified \\
3612227169936143360 & 207.46687 & -13.22687 & 13.608 & 1.959 & 0.150764 &   EBCV \\
2233452336170027648 & 355.08607 & 76.70291 & 17.195 & 0.925 & 0.1546 &   CV \\
2824150286583562496 & 350.78528 & 18.4163 & 14.673 & 1.369 & 0.158263 &   CV \\
405873692214746368 & 24.28709 & 50.95565 & 14.076 & 0.736 & 0.161793 &   CV \\
1780220380340013568 & 328.23357 & 19.28299 & 16.887 & 0.656 & 0.162965 &   EB \\
518137368471752960 & 30.2262 & 64.94512 & 14.174 & 1.698 & 0.168179 &   Unclassified \\
2843374388402149632 & 344.70122 & 25.26188 & 13.462 & 0.607 & 0.17368 &   EBCV \\
6179636986011888512 & 198.32121 & -32.98694 & 14.967 & 1.46 & 0.174586 &   CV \\
2528989654282071936 & 11.65842 & -3.65236 & 16.24 & 1.58 & 0.175009 &   EB \\
635508894697188864 & 137.46053 & 18.82977 & 15.975 & 1.459 & 0.175467 &   EBCV \\
674214551557961984 & 118.77167 & 22.00122 & 13.903 & 1.267 & 0.176907 &   CV \\
4676468869876208128 & 65.16568 & -62.75021 & 17.385 & 0.981 & 0.177177 &   Unclassified \\
2250058535161610240 & 301.41831 & 68.54072 & 16.481 & 0.606 & 0.187861 &   EB \\
5307787074808526592 & 146.0391 & -56.28656 & 16.805 & 0.802 & 0.187952 &   CV \\
2262915708740112000 & 284.3349 & 71.52188 & 16.228 & 0.839 & 0.189128 &   CV \\
5985406470331309312 & 235.21538 & -49.95193 & 16.527 & 0.797 & 0.207472 &   Unclassified \\
2258357545848389888 & 271.05977 & 67.90346 & 14.415 & 1.131 & 0.209913 &   CV \\
6352580849275427072 & 356.04797 & -79.49416 & 16.116 & 0.753 & 0.210131 &   Unclassified \\
4658696016034748928 & 80.40076 & -68.47198 & 16.474 & 1.562 & 0.211071 &   Unclassified \\
6467622480308296448 & 306.43323 & -58.39283 & 16.341 & 0.813 & 0.221345 &   Unclassified \\
3582506919665391616 & 182.7538 & -7.87818 & 15.882 & 0.21 & 0.221523 &   EB \\
1187117793700269952 & 224.14262 & 16.1938 & 17.817 & 0.074 & 0.229145 &   EB \\
3212985152042613120 & 76.15245 & -3.63502 & 13.539 & 1.866 & 0.239897 &   Unclassified \\
5439392500602380544 & 145.71882 & -32.97526 & 15.331 & 0.953 & 0.242473 &   Unclassified \\
4661744416353375488 & 72.70446 & -67.03681 & 13.287 & 1.293 & 0.257954 &   Unclassified \\
4858226697521522432 & 60.50269 & -34.74394 & 16.304 & 1.326 & 0.272607 &   Unclassified \\
4896134903510805888 & 66.64027 & -25.04119 & 14.979 & 1.248 & 0.275918 &   Unclassified \\
5237255534212898816 & 172.60475 & -65.26955 & 16.261 & 0.898 & 0.27992 &   Unclassified \\
15207693216816512 & 47.72934 & 9.82344 & 15.158 & 0.263 & 0.286717 &   CV \\
4788741548375134336 & 65.27338 & -48.65433 & 13.644 & 1.866 & 0.303676 &   EB \\
5216574579445522304 & 132.9392 & -74.666 & 16.875 & 1.759 & 0.330013 &   Unclassified \\
2584756467429594880 & 17.53811 & 13.43775 & 16.706 & -0.069 & 0.332675 &   EB \\
3697668912861183872 & 183.24264 & -1.38609 & 16.666 & 0.635 & 0.335848 &   EB \\
2451154710754640640 & 20.77074 & -17.84431 & 15.256 & 1.775 & 0.346029 &   Unclassified \\
1255522363433681792 & 215.97938 & 24.15678 & 17.705 & 0.097 & 0.381991 &   Unclassified \\
5483999652978511232 & 102.61762 & -57.7979 & 16.934 & 1.752 & 0.465164 &   EB \\
1006621281985546240 & 98.37658 & 61.39074 & 14.305 & 1.718 & 0.632075 &   Unclassified \\
2919098716285083776 & 103.72095 & -27.20218 & 17.177 & 1.751 & 0.662612 &   Unclassified \\
1548104507825815296 & 183.93308 & 52.51644 & 12.571 & 1.808 & 0.6676 &   CV \\
6509224907926516480 & 331.02602 & -54.03593 & 16.739 & 1.921 & 0.828164 &   Unclassified \\
1600249121650105472 & 226.52253 & 54.47185 & 15.968 & 1.112 & 0.931269 &   EB \\
2937111534244417664 & 97.85957 & -21.93285 & 16.381 & 2.058 & 1.295823 &   Unclassified \\
418413313096994560 & 11.46564 & 55.21016 & 16.984 & 0.033 & 1.3901 &   Unclassified \\
2931274020848547200 & 108.38822 & -18.40395 & 16.387 & 1.18 & 1.499116 &   Unclassified \\
5375241185442342144 & 173.2349 & -46.65475 & 16.596 & 1.076 & 1.567165 &   Unclassified \\
5337884487666708096 & 167.44249 & -59.16384 & 16.179 & 1.874 & 1.663699 &   Unclassified \\
4677301303256865152 & 67.54627 & -62.14514 & 16.936 & 1.842 & 2.045161 &   Unclassified \\
6130145940426733824 & 185.85461 & -48.24823 & 16.217 & 1.012 & 2.330113 &   Unclassified \\
5516175520745007744 & 122.90484 & -47.97794 & 16.472 & 1.473 & 3.826666 &   Unclassified \\
4298332012051652352 & 299.1892 & 8.57205 & 17.757 & 0.11 & 6.070805 &   Unclassified \\
\enddata
\end{deluxetable}
\clearpage

\bibliography{Main}{}
\bibliographystyle{aasjournal}

\begin{figure}
    \centering
    \includegraphics[width=1.0\linewidth]{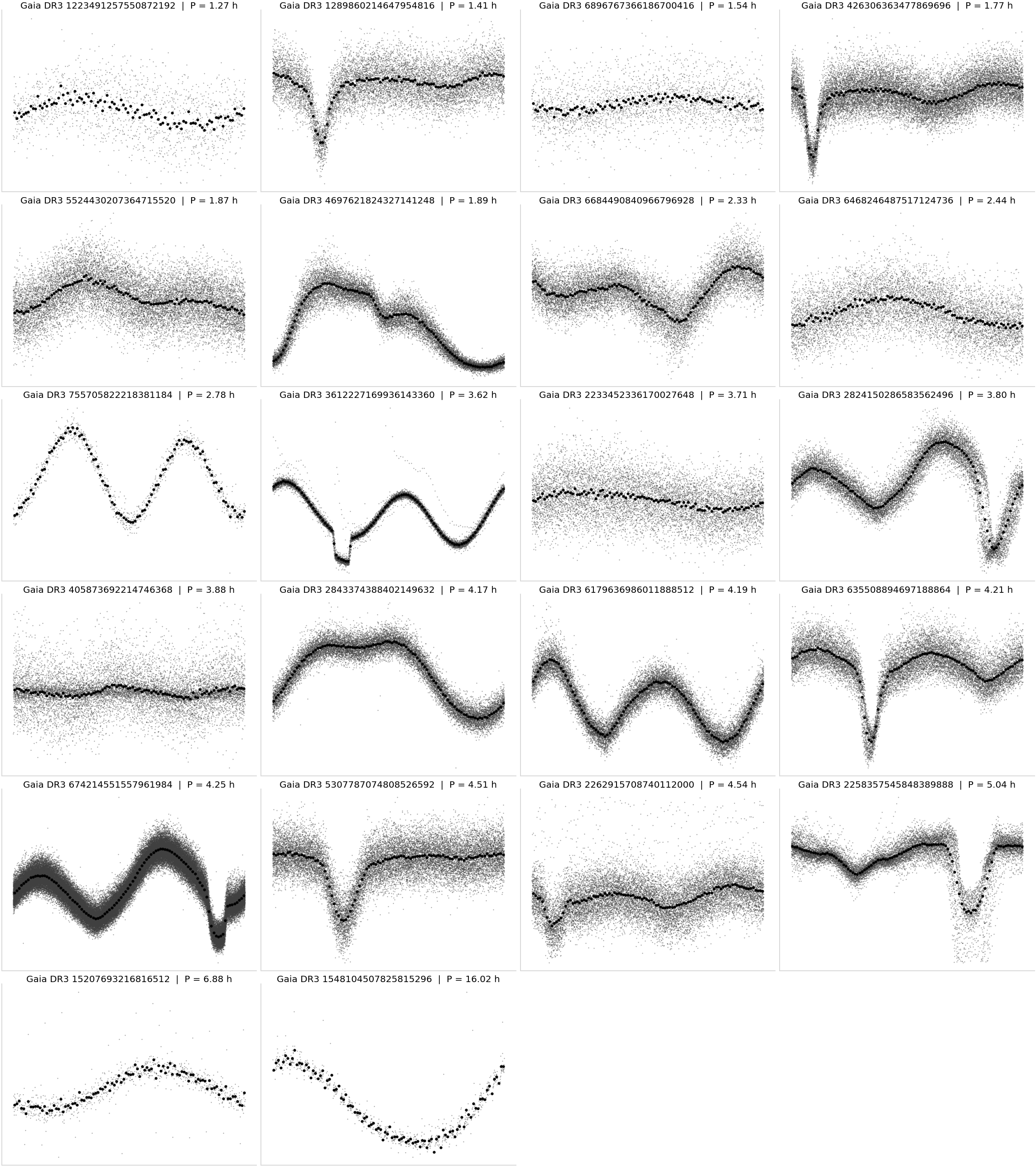}
    \caption{Phase-folded (gray) and binned (black) TESS light curves of the known CVs in our sample.}
    \label{fig:ecl1}
\end{figure}

\begin{figure}
    \centering
    \includegraphics[width=1.0\linewidth]{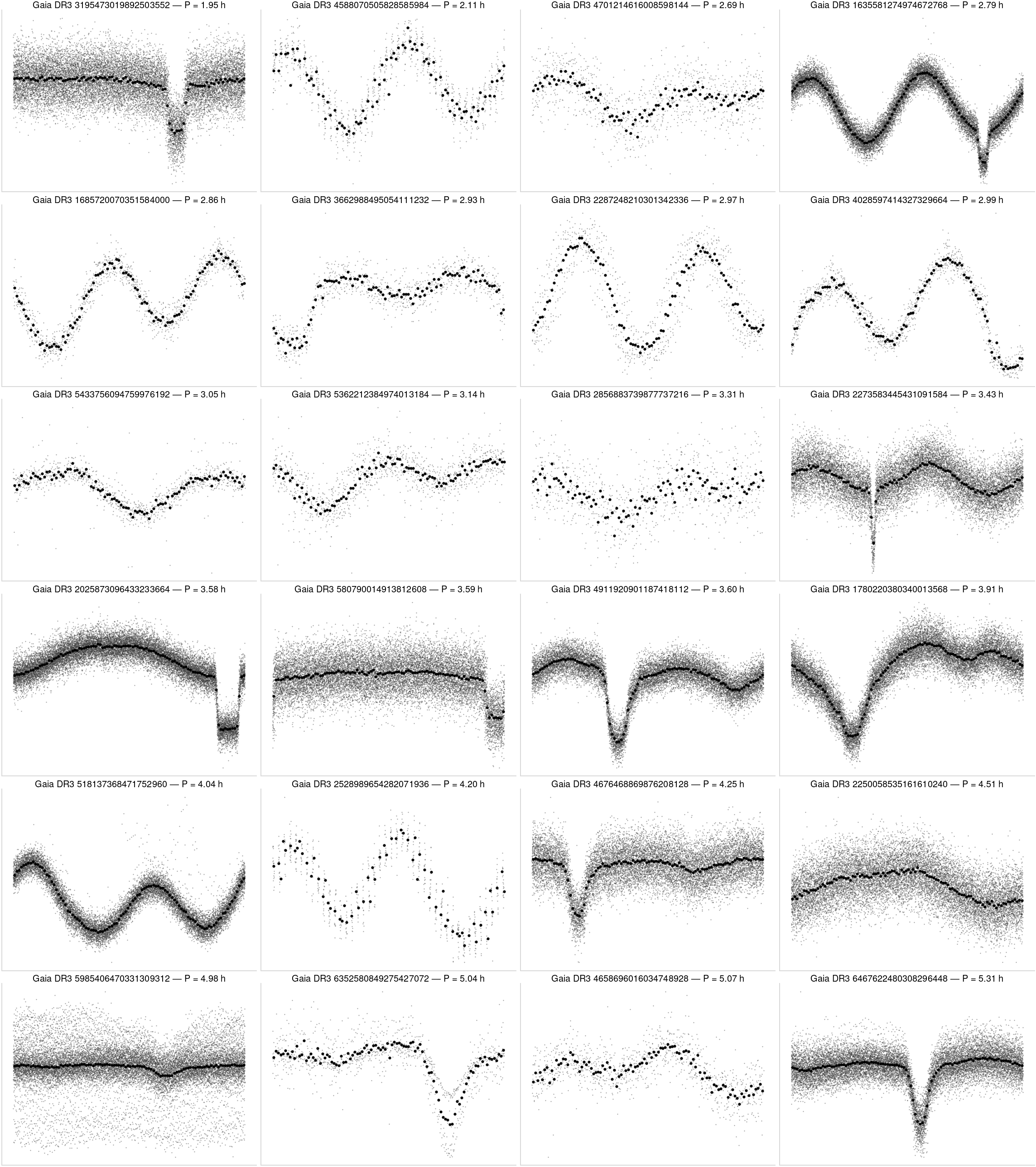}
    \caption{Same as Figure~\ref{fig:ecl1}, but for the EBs and unclassified eclipsing systems in our sample.}
    \label{fig:ecl2}
\end{figure}
\begin{figure}
    \ContinuedFloat
    \centering
    \includegraphics[width=1.0\linewidth]{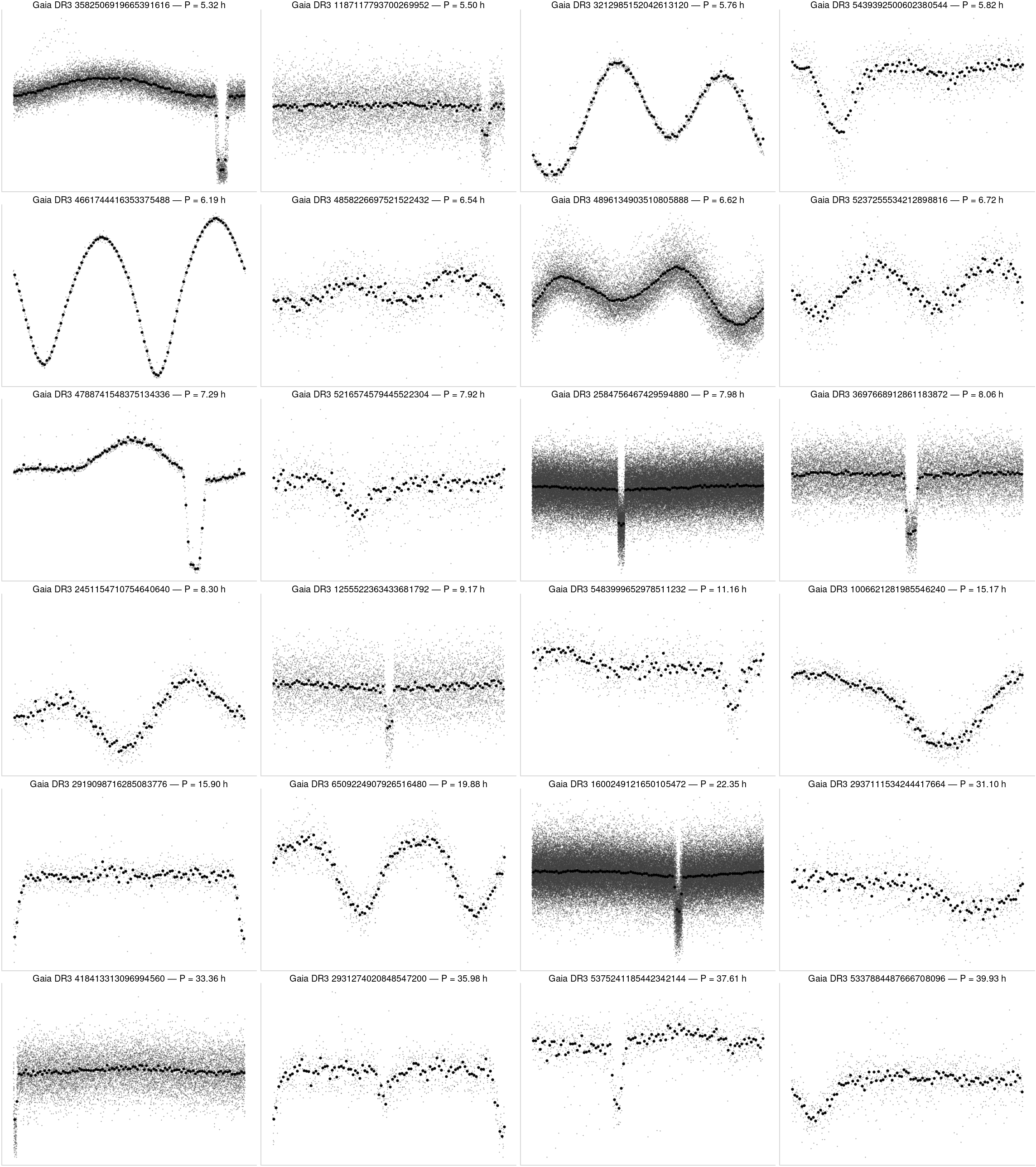}
    \caption{Same as Figure~\ref{fig:ecl1}, but for the EBs and unclassified eclipsing systems in our sample (continued).}
    \label{fig:ecl2l}
\end{figure}
\begin{figure}
    \ContinuedFloat
    \centering
    \includegraphics[width=1.0\linewidth]{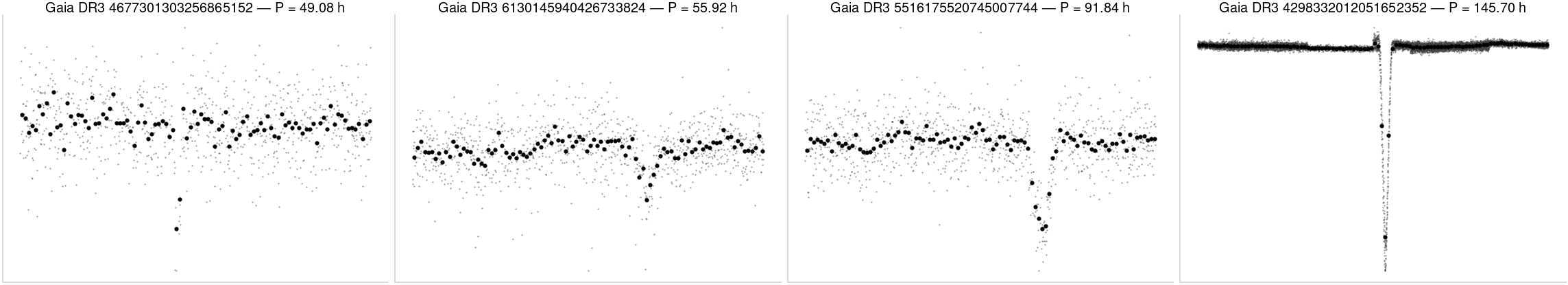}
    \caption{Same as Figure~\ref{fig:ecl1}, but for the EBs and unclassified eclipsing systems in our sample (continued).}
    \label{fig:ecl3}
\end{figure}



\begin{figure}
    \centering
    \includegraphics[width=1.0\linewidth]{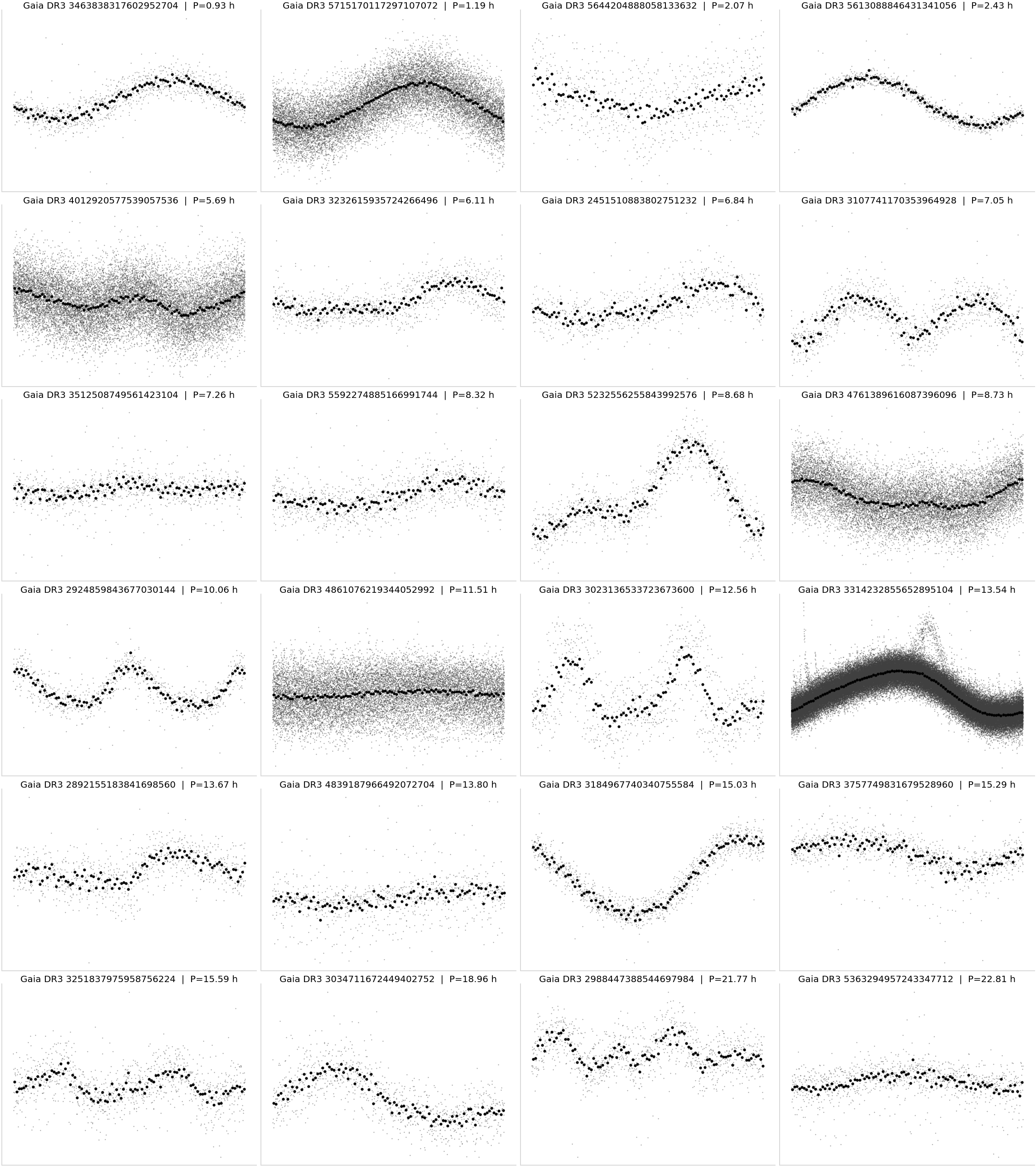}
    \caption{Same as Figure~\ref{fig:ecl1}, but for the systems with sinusoidal variations in our sample.}
    \label{fig:sinusoidal}
\end{figure}

\begin{figure}
    \ContinuedFloat
    \centering
    \includegraphics[width=1.0\linewidth]{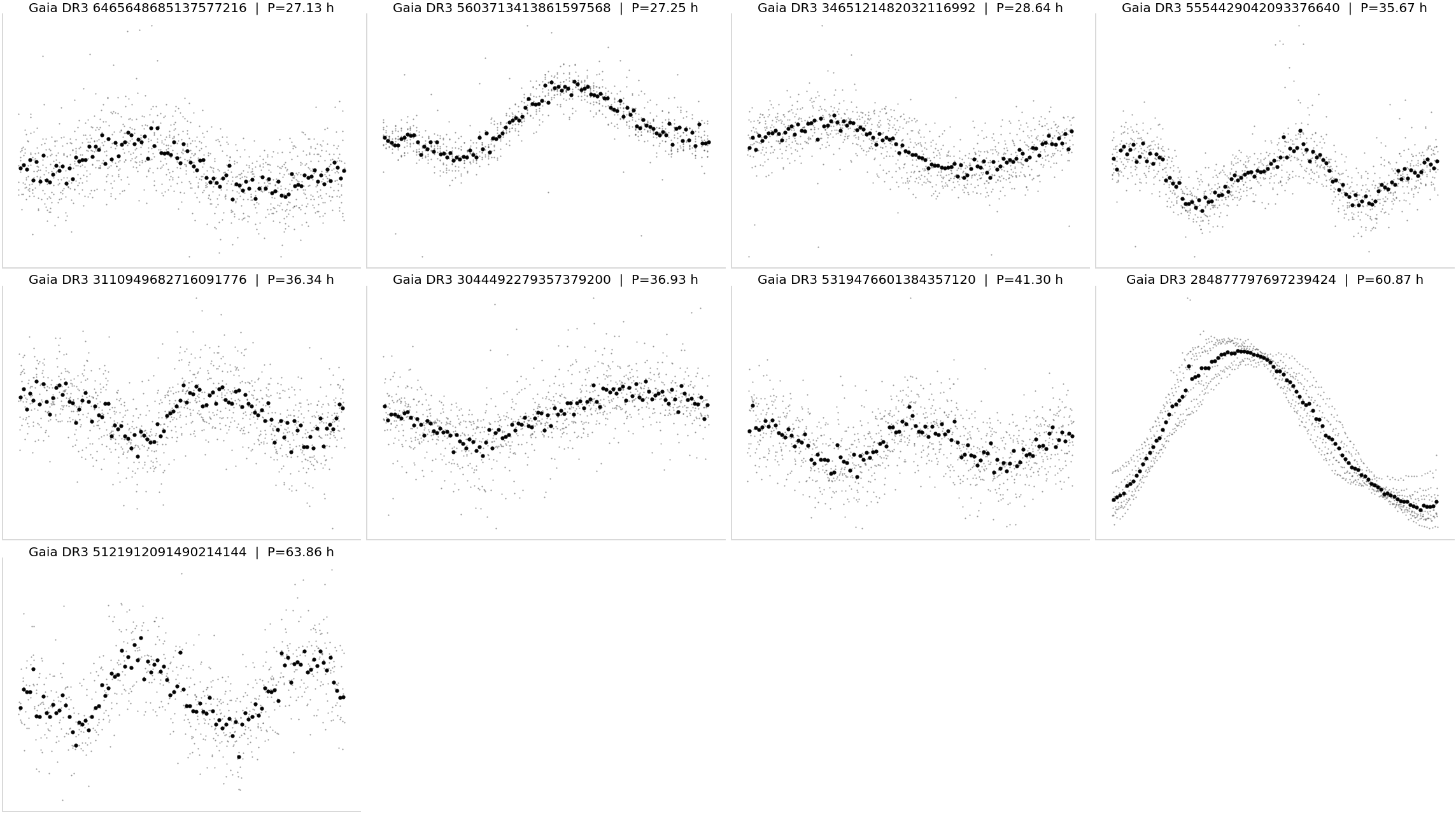}
    \caption{Same as Figure~\ref{fig:ecl1}, but for the systems with sinusoidal variations in our sample (continued).}
\end{figure}



\end{document}